\pdfoutput=1
\documentclass[twocolumn]{autart}
\usepackage[authoryear]{natbib}

\usepackage{graphics}
\usepackage{graphicx}
\usepackage{epsfig}
\usepackage{color}

\usepackage{subfig}
\usepackage{amssymb}
\usepackage{amsmath}
\usepackage{amsfonts}
\usepackage{stmaryrd}
\usepackage{txfonts}
\usepackage{mathrsfs}
\usepackage{latexsym, bm}
\usepackage{xcolor,stfloats}
\usepackage{lipsum}%
\usepackage{indentfirst}
\usepackage{amsmath,amssymb,bm,bbm}

\usepackage{latexsym}
\usepackage{amssymb}
\usepackage{pdfpages}
\def\QEDclosed{\mbox{\rule[0pt]{1.3ex}{1.3ex}}} 

\def\QED{\QEDclosed} 
\def\pf{\noindent{\bf Proof. }}
\def\endpf{\hspace*{\fill}~\QED\par\endtrivlist\unskip}

\newtheorem{theorem}{Theorem}
\newtheorem{corollary}{Corollary}
\newtheorem{lemma}{Lemma}
\newtheorem{definition}{Definition}
\newtheorem{remark}{Remark}
\newtheorem{example}{Example}
\newtheorem{property}{Property}
\newtheorem{assumption}{Assumption}

\begin{document}

\begin{frontmatter}

\title{Opinion dynamics in social networks with stubborn agents: An issue-based perspective \thanksref{label1}}
\thanks[label1]{This work was supported by National Science Foundation of China (Grant Nos. 61533001 and 61375120) and the Fundamental Research Funds for the Central University (Grant No. JBZ170401).}

\author[xidian]{Ye Tian}, \ead{tinybeta7.1@gmail.com}
\author[Peking]{Long Wang\corauthref{cor1}} \ead{longwang@pku.edu.cn}
\corauth[cor1]{Corresponding author : Long Wang }

\address[xidian]{ Center for Complex Systems, School of Mechano-electronic Engineering, Xidian University, Xi'an 710071, P.~R.~China}
\address[Peking]{Center for Systems and Control, College of Engineering, Peking University, Beijing 100871, P.~R.~China}

\begin{abstract}
Classic models on opinion dynamics usually focus on a group of agents forming their opinions interactively over a single issue. Yet generally agreement can not be achieved over a single issue when agents are not completely open to interpersonal influence. In this paper, opinion formation in social networks with stubborn agents is considered over issue sequences. The social network with stubborn agents is described by the Friedkin-Johnsen (F-J) model where agents are stubborn to their initial opinions. Firstly, we propose a sufficient and necessary condition in terms of network topology for convergence of the F-J model over a single issue. Secondly, opinion formation of the F-J model is investigated over issue sequences. Our analysis establishes connections between the interpersonal influence network and the network describing the relationship of agents' initial opinions for successive issues. Taking advantage of these connections, we derive the sufficient and necessary condition for the F-J model to achieve opinion consensus and form clusters over issue sequences, respectively. Finally, we consider a more general scenario where each agent has bounded confidence in forming its initial opinion. By analyzing the evolution of agents' ultimate opinions for each issue over issue sequences, we prove that the connectivity of the state-dependent network is preserved in this setting. Then the conditions for agents to achieve opinion consensus over issue sequences are established. Simulation examples are provided to illustrate the effectiveness of our theoretical results.
\end{abstract}

\end{frontmatter}
\section{\bf Introduction}\label{s-introduction}

Recently, opinion dynamics has attracted much attention of researchers from various disciplines, such as applied mathematics, economics, social psychology, control theory, etc., due to its broad applications in modeling and explaining complex phenomena in social and artificial networks \citep{Degroot74, Lamport82, Krause, Acemoglu10, Frasca15, Ravazzi15}. In a social network, agents form opinions on various political, economic and social issues according to the information they received from neighbors determined by the network topology \citep{jadbabaie03, sabertac04, ren05} or the confidence/influence bound \citep{Blondel09, Bullo12, J14}. A fundamental question in opinion dynamics is: how do the network structure and opinions' initial distribution influence the diffusion and aggregation of scattered opinions in the process of opinion formation? In \cite{Degroot74}, a model is presented to characterize the process of a group of agents reaching opinion consensus on a common issue by pooling their subjective opinions, which is known as the Degroot model. The interactions between agents are described by a stochastic matrix which can be regarded as the one-step transition probability matrix of a Markov chain and some sufficient conditions for achieving opinion consensus are provided. To further investigate how the interpersonal influence contributes to the opinion formation, the work in \cite{Friedkin99} extends the Degroot model by introducing a diagonal matrix which represents agents' susceptibilities to interpersonal influence. In the Degroot model and the Friedkin-Johnsen (F-J) model, interactions between agents are specified by given networks. Different from these linear models, in \cite{Krause}, the authors present a nonlinear model in which agents have bounded confidence for others. In the Hegselmann-Krause (H-K) model, two agents are said to be connected if and only if the difference between their opinions is smaller than a given confidence bound, which means that the network topology of the H-K model is state-dependent.

As the research in multi-agent systems continues, the tools that are available within opinion dynamics have been enriched considerably. The Degroot model is further investigated both on continuous-time dynamics and switched topologies \citep{jadbabaie03, sabertac04, ren05}. In addition, many complex scenarios are also considered, including asynchronous consensus \citep{Xiao08-1}, time delays \citep{Wang07}, finite-time consensus \citep{Wang10}, leader-following framework \citep{Liu12, Ma16}, heterogeneous model \citep{Zheng14} and antagonistic interactions \citep{Altafini13}, to name but a few. Based on the gossip algorithm \citep{Boyd06}, randomized opinion dynamics is investigated in \cite{Acemoglu10}, \cite{Ravazzi15}, \cite{Frasca15} and \cite{Li13}, respectively. In \cite{Javad14}, the equilibrium and convergence rate of the F-J model are investigated. Under the assumption that the adjacent matrix of the interpersonal influence network is an irreducible sub-stochastic matrix with the row sum of at least one row strictly smaller than one, the authors transform the F-J model to a random walk, and the form of equilibrium is proposed based on the first hitting probabilities of the random walk. In \cite{Bindel11}, the F-J model is interpreted as a best-response game, and the ratio between the optimal solution and the Nash equilibrium solution, which is defined as the price of anarchy, is discussed under both undirected and directed networks.

Most of the available literature on the F-J model focuses on opinion evolving over a single issue. Yet the F-J model does not in general converge to consensus over a single issue due to the presence of stubborn agents. Actually, the empirical evidence shows that consensus may be reached over a sequence of issues \citep{Joshi10}. In practice, associations of individuals (such as small groups within firms, deliberative bodies of government, etc.) are usually constituted to deal with issues within particular issue domains which consist of deeply interdependent issues, especially repeatedly arising issues. In this scenario, individuals' opinions for interdependent issues are always correlated. Thus, extending the existing theories of opinion dynamics to issue sequences is necessary and can uncover the underlying mechanism of opinion formation in the real world. \cite{Bullo14} and \cite{Bullo15} modify the F-J model and the Degroot model to investigate the evolution of agents' self-appraisals over an issue sequence, respectively. In \cite{Tempo15}, the authors present a multidimensional extension of the F-J model in which agents' opinions for several interdependent issues evolve over time sequences.

Different from the works which focus on agents' self-appraisal dynamics \citep{Bullo14, Bullo15} over sequences of independent issues, or evolution of agents' opinions for multiple interdependent but unordered issues over time sequences \citep{Tempo15}, we consider the opinion formation of the F-J model which evolves over both interdependent issue sequences and time sequences in this paper. Firstly, convergence of the F-J model is studied over a single issue. The existing results about convergence of the F-J model over a single issue usually require that the interpersonal influence network is undirected and connected or its adjacent matrix is strictly row-substochastic \citep{Javad14, Bullo14}. We propose a milder sufficient and necessary condition in terms of network topology to guarantee that opinions of agents converge to constant values. Several properties of agents' ultimate opinions for a single issue are provided. Secondly, we study opinion formation of the F-J model over a sequence of interdependent issues inspired by the path-dependence theory \citep{North90, Page06, MacKay06, Egidi97}. By virtue of the inherent coherence between basic assumption of the F-J model and the path-dependence theory, we assume that the factor of each agent's cognitive inertia over two interdependent issues equals to its stubborn factor. Then, each agent will form its initial opinion for the next issue by making a tradeoff between its initial opinion for the last interdependent issue and other agents' initial opinions for the next issue. The connections between the interpersonal influence network and the network which characterizes the relationship of agents' initial opinions for successive issues are established, and the sufficient and necessary condition for the F-J model to achieve opinion consensus or form clusters over issue sequences is proposed. Finally, we consider the more general case in which an information assimilation mechanism is employed to weaken agents' interpersonal influence. We assume that each agent maintains a confidence bound in forming its initial opinions. This assumption is consistent with the reported echo-chamber effect \citep{Sunstein12, Dandekar13}, i.e., people usually assimilate information in a selective way: they tend to give considerable weight to the information supporting their initial opinions, and dismiss the undermining information meanwhile. Connectivity preservation of the modified F-J model is analyzed. Then we derive the conditions for achieving opinion consensus. The main difficulties for our analysis are twofold. On one hand, the evolution of agents' initial opinions over issue sequences is not directly determined by the interpersonal influence network and agents' stubborn extent, which increases complexity of our analysis from the graphical perspective. On the other hand, due to the fact that the evolution of agents' initial opinions over issue sequences is influenced by the evolution of their opinions over time sequences, connectivity preservation of the modified F-J model is more complex compared with connectivity preservation of the H-K model. As has been widely reported \citep{Javad14, Frasca15, Friedkin15}, a social network with stubborn agents does not generally achieve consensus over a single issue. Our investigation shows that opinions of a group of individuals can achieve consensus over each path-dependent issue sequence \footnote{In the sequel, we say that an issue sequence is path-dependent or has the property of path-dependence if individual's opinions over it have the path-dependent property. A detailed explanation of path dependence will be presented below.}, even if agents are stubborn to their initial opinions. Simply put, repeatedly arising or interdependent issues lead to consensus. Our study also provides a theoretical explanation for the cohort effect \citep{Joshi10}. Moreover, the evolution of opinions over issue sequences enhances the connectivity of the social network. As long as one of the partially stubborn agent's opinions can scatter to all the rest partially stubborn agents through partially stubborn or non-stubborn agents, opinion consensus can be reached, and all the partially stubborn agents form a star subgraph in the network which characterizes the relationship of agents' initial opinions for successive issues. The differences between opinion dynamics over issue sequences and a single issue give us a deeper understanding of opinion formation in social networks.

The rest of this paper is organized as follows. In Section \ref{s-Preliminary}, we introduce some notions on graph theory and the F-J model. In Section \ref{s-convergence}, convergence of the F-J model is studied over a single issue, and some properties of ultimate opinions are proposed. In Section \ref{s-FJs}, opinion formation of the F-J model is investigated over issue sequences. In Section \ref{s-FJh}, we consider opinion formation of the F-J model over issue sequences with bounded confidence. In Section \ref{s-Simulations}, numerical simulations are given to illustrate the effectiveness of theoretical results. Some conclusions are drawn in Section \ref{s-Conclusion}.

\textbf{Notation}: Denote the set of real numbers, natural numbers, $n$-dimensional real vector space and $n\times n$ real square matrix space by $\mathbb{R}$, $\mathbb{N}$, $\mathbb{R}^{n}$ and $\mathbb{R}^{n\times n}$, respectively. $\mathbf{1}_{n}$ is a vector of size $n$ having all the elements equal $1$. $I_{n}$ is the $n\times n$ identity matrix. For a given vector or matrix $A$, $A^{T}$ denotes its transpose, $\|A\|_{\infty}$ denotes its maximum row sum norm. $0$ denotes an all-zero vector or matrix with compatible dimension. $\rho(A)$ denotes the spectral radius for a matrix $A\in\mathbb{R}^{n\times n}$. By $|\cdot|$ we denote the absolute value, modulus and cardinality of real number, complex number and set, respectively. Given a vector $\zeta\in\mathbb{R}^{n}$ and a matrix sequence $\{A_{k}\}_{k=1}^{m}$ with $A_{k}\in\mathbb{R}^{n\times n}$, $diag(\zeta)$ denotes an $n\times n$ diagonal matrix whose diagonal elements are the elements of $\zeta$, $diag(A_{1}, A_{2},\ldots,A_{m})$ denotes an $nm\times nm$ block diagonal matrix with $A_{k}$ being its block on diagonal.


\section{\bf Preliminaries}\label{s-Preliminary}

In this section, some basic concepts on graph theory are introduced. Then, a brief introduction of the F-J model is presented.

\subsection{Basic concepts on graph theory}\label{s-graph theory}

A weighted directed graph (digraph) $\mathcal{G}$ of order $n$ is a triple $\mathcal{G}(W)=(V,E,W)$ which consists of a vertex set $V=\{1, 2, \ldots, n\}$, an edge set $E=\{e_{ij}:i, j\in V\}$ and a weighted adjacency matrix $W=[w_{ij}]_{n\times n}$ with entries $w_{ij}$. An edge $e_{ij}$ is an ordered pair $(i, j)$ which means that node \emph{j} can obtain information from node \emph{i}, but not necessarily vice versa. $i$ is called the parent of $j$ if $(i,j)\in E$. The adjacency elements associated with the edges of the graph are nonzero, i.e., $e_{ij}\in E$ means $w_{ji}\neq0$, otherwise, $w_{ji}=0$. A directed path of length $m+1$ from $i$ to $j$ is a finite ordered sequence of distinct vertices of $\mathcal{G}$ with the form $i, k_{1}, k_{2}, \cdots, k_{m}, j$. Particularly, if $i=j$, the directed path is called a cycle of $i (j)$. The directed graph $\mathcal{G}$ is said to be strongly connected if between every pair of distinct vertices $i,j$, there exists a directed path that begins at $i$ and ends at $j$. A maximal subgraph of $\mathcal{G}$ that is strongly connected forms a strongly connected component (SCC). In an SCC of digraph $\mathcal{G}$, if there exists a vertex which has parents belonging to other SCCs, we say that the in-degree of this SCC is nonzero, otherwise, we call it an independent strongly connected component (ISCC). The directed graph $\mathcal{G}$ is said to have a spanning tree if there exists a vertex that is called the root which has a directed path to every other vertex. The directed graph $\mathcal{G}$ is said to contain a star subgraph if there exists a vertex that is called the center vertex which has directed edges to any other vertices. An undirected graph is a digraph $\mathcal{G}(W)$ which satisfies that $w_{ij}=w_{ji}$ for any $i, j\in V$. For more details about algebraic graph theory, we refer to \cite{Godsil01}.

\subsection{Introduction of the F-J model}\label{s-System model}

The F-J model presented in \cite{Friedkin99} sheds light on the relationship between interpersonal influence and opinion change in social networks. The F-J model is a generalization of the Degroot model \citep{Degroot74}. It introduces a positive diagonal matrix to quantify the extent of each member of the group open to the interpersonal influence.

Consider $n$ agents forming opinions on a certain issue in a network formulated by a weighted digraph $\mathcal{G}(W)=(V,E,W)$. Each agent holds an initial opinion $x_{i}(0), i\in V$. The classic F-J model is:
\begin{equation}\label{FJ}
 x(k+1)=\Xi Wx(k)+(I-\Xi)x(0),
\end{equation}
where $x(k)=(x_{1}(k),x_{2}(k),\ldots,x_{n}(k))^{T}\in\mathbb{R}^{n}$ is a vector representing the stack of opinions with $k\in\mathbb{N}$ being the scale of time, $W$ is a stochastic weighted adjacency matrix characterizing the structure of the interpersonal influence network, $\Xi=diag(\xi)$ is the diagonal matrix mentioned above with $\xi=(\xi_{1},\xi_{2},\ldots,\xi_{n})^{T}\in\mathbb{R}^{n}$ satisfying $\xi_{i}\in[0,1]$. $\xi_{i}$ represents the susceptibility of agent $i$ to interpersonal influence while $1-\xi_{i}$ quantifies the extent to which agent $i$ is stubborn to its initial opinion. Here, $x_{i}(0)$ can be regarded as the internal opinions of agent $i$ for the issue \citep{Bindel11}. Note that if $\xi=\mathbf{1}_{n}$, the F-J model collapses to the Degroot model. However, different from the Degroot model in which disagreement is mainly the consequence of lack of communication, it is more likely the result of the presence of stubborn agents anchored to their own initial opinions in the F-J model. This is coincident with the empirical evidence that failing in reaching agreement usually owes to the stubbornness of individuals in practice, rather than the insufficiency of persistent contracts and interactions \citep{Friedkin99, Friedkin11, Frasca15}.

As mentioned above, if $\Xi=I$ in equation (\ref{FJ}), the F-J model is transformed into the Degroot model:
\begin{equation}\label{De}
  x(k+1)=Wx(k),
\end{equation}
where $W$ is a stochastic matrix. On the other hand, consider the augmented system of the F-J model (\ref{FJ}). Let $\hat{x}(k)=[x(0)^{T} x(k)^{T}]^{T}$, we have
\begin{equation}\label{aug1}
 \hat{x}(k+1)=\hat{W}\hat{x}(k),
\end{equation}
where
\begin{equation*}
 \hat{W}=\begin{bmatrix}
  I_{n\times n} & 0 \\
  (I-\Xi) & \ \ \ \Xi W
  \end{bmatrix}.
\end{equation*}
From equation (\ref{aug1}), the F-J model is equivalent to the leader-following structure \citep{Liu12} where agents' initial opinion are virtualized as stationary leaders.

\section{Convergence of the F-J model over a single issue}\label{s-convergence}

A fundamental problem with respect to the F-J model is its limiting behavior when the time scale $k$ tends to infinity. In this subsection, convergence of the F-J model is considered over a single issue. A sufficient and necessary condition on the regularity of $\Xi W$ for convergence of the F-J model is presented in \cite{Tempo15}. Here we give a sufficient and necessary condition from the perspective of the interpersonal influence network $\mathcal{G}(W)$ which ensure that $x(k)$ converges to a constant vector as $k\rightarrow\infty$. Firstly, we propose the definition of the convergence of dynamic systems.

\begin{definition}\label{D1}
For a dynamic system with state vector $\ell(k)\in\mathbb{R}^{n}, k\in\mathbb{N}$, if for any initial state $\ell(0)$, there exists a constant vector $\hat{\ell}\in\mathbb{R}^{n}$, such taht
\begin{equation*}
 \lim_{k\rightarrow\infty}\ell(k)=\hat{\ell},
\end{equation*}
we say the system is convergent.
\end{definition}

\begin{lemma}\label{L1}
The F-J model (\ref{FJ}) is convergent if and only if $1$ is the only maximum-modulus eigenvalue of $\Xi W$.
\end{lemma}

Lemma \ref{L1} naturally follows from Corollary 2 and Lemma 5 in \cite{Tempo15}. In Lemma \ref{L1}, that $1$ is the only maximum-modulus eigenvalue of $\Xi W$ means $\Xi W$ has no other maximum-modulus eigenvalues except $1$, here the algebraic multiplicity of eigenvalue $1$ may be larger than one. In other words, suppose $\lambda$ is an eigenvalue of $\Xi W$, then $|\lambda|=1$ implies $\lambda=1$. Otherwise, let $a+bi\neq1$ be an eigenvalue of $\Xi W$ with $|a+bi|=1$. We have $a+bi=e^{i\arctan\frac{b}{a}}=e^{i\theta}$, where $\theta\in[-\pi,\pi]$ is the principal value of the argument of $a+bi$. Thus, $(a+bi)^{k}=e^{i\theta k}$, which shows that $(a+bi)^{k}$ is periodic. Suppose that $J$ is the Jordan canonical form of $\Xi W$, then $J^{k}$ is periodic, which indicates that system (\ref{FJ}) is non-convergent. Let $\Xi=I$, one can readily get that the Degroot model is convergent if and only if $1$ is the only maximum-modulus eigenvalue of $W$.

For two positive integers, we say they are coprime if and only if their greatest common divisor is $1$. For agent $i$, if $\xi_{i}<1$, we say it is a stubborn agent, otherwise, we say it is a non-stubborn agent. In \cite{Golub10}, the authors prove that the Degroot model (\ref{De}) is convergent if and only if each ISCC of $\mathcal{G}(W)$ is aperiodic from the perspective of Markov chain corresponding to $W$. In \cite{Friedkin15}, convergence of the F-J model requires that the lengths of all cycles of $\mathcal{G}(\Xi W)$ are coprime. In the next theorem, we propose a graph-theoretical condition for the F-J model to achieve convergence by virtue of properties of irreducible and primitive matrix's eigenvalue. Different from the Degroot model (\ref{De}), convergence of the F-J model only depends on the ISCC composed of non-stubborn agents in $\mathcal{G}(W)$. Before proposing our convergence condition, we need the following lemmas in \cite{Horn85}.

\begin{lemma}\label{L1.1}
Let $A\in\mathbb{R}^{n\times n}$ be irreducible and suppose that $\lambda\in\mathbb{R}$ is not in the interior of any $Ger\breve{s}gorin$ disc of $A$. If some $Ger\breve{s}gorin$ circle of A does not pass through $\lambda$, then $\lambda$ is not an eigenvalue of $A$.
\end{lemma}

\begin{lemma}\label{L1.2}
Let $A\in\mathbb{R}^{n\times n}$ be irreducible and nonnegative, and let $P_{1}, P_{2},\ldots, P_{n}$ be the vertices of the directed graph $\mathcal{G}(A)$. Let $L_{i}=\{l^{i}_{1}, l^{i}_{2},\ldots \}$ be the set of lengths of all cycles of $P_{i}$ in $\mathcal{G}(A)$, Let $\gamma_{i}$ be the greatest common divisor of all the lengths in $L_{i}$. Then A is primitive if and only if $\gamma_{1}=\gamma_{2}=\cdots=\gamma_{n}=1$.
\end{lemma}
\begin{remark}
Note that the condition in Lemma \ref{L1.2} is consistent with the definition of aperiodicity (Definition 2 in \cite{Golub10}). For a nonnegative and irreducible matrix, aperiodicity ensures that it does not have any periodic eigenvalues, that is , it is primitive. Thus, aperiodicity is a graph-theoretical criterion for primitivity. By the virtue of aperiodicity, we provide a sufficient and necessary condition concerning on the interpersonal influence network $\mathcal{G}(W)$ for the F-J model to achieve convergence in the next theorem.
\end{remark}
\begin{assumption}\label{A1}
For each ISCC of $\mathcal{G}(W)$ which only consists of more than one non-stubborn agent, there exists at least one agent having two cycles which are coprime in length.
\end{assumption}
Assumption \ref{A1} means that all ISCCs of $\mathcal{G}(W)$ consisting of only non-stubborn agents are aperiodic. This condition is milder than condition for convergence of the Degoot model (\ref{De}) which requires that all ISCC of $\mathcal{G}(W)$ are aperiodic \citep{Golub10}.
\begin{theorem}\label{T1}
The F-J model (\ref{FJ}) achieves convergence if and only if Assumption \ref{A1} holds.
\end{theorem}
{\bf Proof.} Suppose that $\mathcal{G}(W)$ has $\alpha$ SCCs with $\alpha\leq n$. Because $W$ is reducible, then there exists a permutation matrix $P$ such that
\begin{equation*}
 \tilde{W}=P^{T}WP=\begin{bmatrix}
W_{1} & \times & \cdots & \times \\
&W_{2} & \cdots &\times \\
& & \ddots & \vdots \\
\multicolumn{2}{c}{\raisebox{1.3ex}[0pt]{\Huge0}}
& &W_{\alpha}
\end{bmatrix},
\end{equation*}
where $W_{i}, i\in\{1, 2, \ldots, \alpha\}$ is the irreducible adjacent matrix corresponding to each SCC of $\mathcal{G}(W)$ \citep{Brualdi91}.

Let $\tilde{\Xi}=P^{T}\Xi P=diag(\Xi_{1}, \Xi_{2},\ldots,\Xi_{\alpha})$, then we have $\Xi W=P\tilde{\Xi}\tilde{W}P^{T}$. If there exists $\xi_{j}=0$ which is a diagonal element of $\Xi_{i}$, then $\mathcal{G}(W_{i})$ is separated into a vertex $j$ which corresponds to a row of $\Xi_{i}W_{i}$ with all elements being zero and several SCCs with nonzero in-degrees in $\mathcal{G}(\Xi_{i}W_{i})$. Otherwise, if $\mathcal{G}(W_{i})$ has nonzero in-degree, or $\mathcal{G}(W_{i})$ is an ISCC but contains at least one stubborn agent, $\Xi_{i}W_{i}$ is a irreducible matrix which satisfies $\|\Xi_{i}W_{i}\|_{\infty}\leq1$ and contains at least one row whose row sum is strictly smaller than $1$. From Lemma \ref{L1.1}, we obtain $\rho(\Xi_{i}W_{i})<1$. If $\mathcal{G}(W_{i})$ is an ISCC and contains no stubborn agent, i.e., $\Xi_{i}W_{i}=W_{i}$ is an irreducible stochastic matrix, then 1 is its only maximum-modulus eigenvalue if and only if $W_{i}$ is primitive. In this case, if $\mathcal{G}(W_{i})$ contains only one non-stubborn agent $i$, then the eigenvalue of $W_{i}$ is $1$. Otherwise, by Lemma \ref{L1.2}, 1 is the only eigenvalue of $W_{i}$ that has maximum modulus if and only if there exists at least one agent in $\mathcal{G}(W_{i})$ which has at least two cycles with coprime length. In conclusion, 1 is the only eigenvalue of $\Xi W$ with maximum-modulus if and only if Assumption 1 holds. By Lemma \ref{L1}, the F-J model (\ref{FJ}) achieves convergence if and only if Assumption \ref{A1} holds. $\blacksquare$

\begin{remark}
In the available literature, convergence of the F-J model over a single issue usually requires that $\mathcal{G}(W)$ is undirected connected or every ISCC of $\mathcal{G}(W)$ contains at least one stubborn agent \citep{Bullo14, Javad14}. Theorem \ref{T1} shows that the convergence is achieved under the milder condition which ensures condition in Lemma \ref{L1}. The proof of Theorem \ref{T1} also implies that each ISCC consisting of non-stubborn agents achieves consensus as the F-J model converges. Note that from Theorem \ref{T1}, that each ISCC consisting of non-stubborn agents of $\mathcal{G}(W)$ has at least one agent containing self-loop is a sufficient but not necessary condition for convergence of the F-J model. However, if $\mathcal{G}(W)$ is undirected, the F-J model achieve convergence if and only if each ISCC consisting of non-stubborn agents has at least one agent containing self-loop in $\mathcal{G}(W)$. The proof directly follows from Theorem \ref{T1}.
\end{remark}

Next, we shall further analyze the ultimate opinion of system (\ref{FJ}). Employing equation (\ref{FJ}) iteratively, we have
\begin{equation}\label{FJ1}
  x(k)=\Psi(k)x(0),
\end{equation}
where
\begin{equation}\label{FJ2}
 \Psi(k)=(\Xi W)^{k}+\sum_{t=0}^{k-1}(\Xi W)^{t}(I-\Xi).
\end{equation}
Since system (\ref{FJ}) is convergent under Assumption \ref{A1}, the limit of $\Psi(k)$ exists. Let $\Psi=\lim_{k\rightarrow\infty}\Psi(k)$.

\begin{property}\label{P1}
$\Psi$ is a stochastic matrix.
\end{property}
{\bf Proof.} Firstly, we show that $\Psi(k)$ is stochastic for any $k\in\mathbb{N}$ inductively. Note that $\Xi W$ and $I-\Xi$ are both nonnegative, thus $\Psi(k)$ is nonnegative. From equation (\ref{FJ2}), we have $\Psi(0)=I$ and $\Psi(1)=\Xi W+I-\Xi$ are both stochastic. Now suppose that $\Psi(l)$ is a stochastic matrix, we have
\begin{equation*}
\begin{split}
  \Psi(l+1)\mathbf{1}_{n}&=((\Xi W)^{l+1}+\sum_{t=0}^{l}(\Xi W)^{t}(I-\Xi))\mathbf{1}_{n}\\
           &=((\Xi W)^{l}\Xi W+\sum_{t=0}^{l-1}(\Xi W)^{t}(I-\Xi)+(\Xi W)^{l}(I-\Xi))\mathbf{1}_{n}\\
           &=((\Xi W)^{l}(\Xi W+I-\Xi)+\sum_{t=0}^{l-1}(\Xi W)^{t}(I-\Xi))\mathbf{1}_{n}\\
           &=(\Xi W)^{l}\Psi(1)\mathbf{1}_{n}+\sum_{t=0}^{l-1}(\Xi W)^{t}(I-\Xi)\mathbf{1}_{n}.
  \end{split}
\end{equation*}
Since $\Psi(1)$ is stochastic, we have $(\Xi W)^{l}\Psi(1)\mathbf{1}_{n}=(\Xi W)^{l}\mathbf{1}_{n}$. Therefore, $\Psi(l+1)\mathbf{1}_{n}=((\Xi W)^{l}\Xi W+\sum_{t=0}^{l-1}(\Xi W)^{t}(I-\Xi))\mathbf{1}_{n}=\Psi(l)\mathbf{1}_{n}=\mathbf{1}_{n}$, which implies that $\Psi(k)$ is stochastic for any $k\in\mathbb{N}$. Moreover, since $\Psi\mathbf{1}_{n}=\lim_{k\rightarrow\infty}\Psi(k)\mathbf{1}_{n}=\mathbf{1}_{n}$, we obtain that $\Psi$ is a stochastic matrix. $\blacksquare$

Property \ref{P1} shows that each agent's opinion converges to a convex combination of all agents' initial opinions. Let $x^{*}=\lim_{k\rightarrow\infty}x(k)$. Denote the set of fully stubborn agents by $V_{f}=\{i\in V|\xi_{i}=0\}$, the set of partially stubborn agents by $V_{p}=\{i\in V|0<\xi_{i}<1\}$ and the set of non-stubborn agents by $V_{n}=\{i\in V|\xi_{i}=1\}$. Without loss of generality, let $V_{f}=\{1, 2, \ldots, r_{1}\}, V_{p}=\{r_{1}+1, r_{1}+2, \ldots, r_{1}+r_{2}\}, V_{n}=\{r_{1}+r_{2}+1, r_{1}+r_{2}+2, \ldots, n\}$. Let $\Psi=[\psi_{1}, \psi_{2}, \ldots, \psi_{n}]$, where $\psi_{i}\in\mathbb{R}^{n}$. The following property reveals the influence of the interpersonal influence network $\mathcal{G}(W)$ and the levels of stubbornness of agents to the ultimate opinion $x^{*}$.

\begin{property}\label{P2}
Under Assumption \ref{A1}, for any $i\in V_{n}$, if there exists $j\in V_{f}\bigcup V_{p}$ which has a directed path to agent $i$ in graph $\mathcal{G}(W)$, then $\psi_{i}=0$.
\end{property}
{\bf Proof.} Following equation (\ref{FJ2}), we obtain
\begin{equation*}
\begin{split}
  &\Psi(k+1)-\Psi(k)\\
   =&(\Xi W)^{k+1}+\sum_{t=0}^{k}(\Xi W)^{t}(I-\Xi)-(\Xi W)^{k}+\sum_{t=0}^{k-1}(\Xi W)^{t}(I-\Xi)\\
   =&(\Xi W)^{k+1}-(\Xi W)^{k}+(\Xi W)^{k}(I-\Xi)\\
   =&(\Xi W)^{k}\Xi (W-I).
  \end{split}
\end{equation*}
From the proof of Theorem \ref{T1}, $\lim_{k\rightarrow\infty}(\Xi W)^{k}$ exists. Moreover, since $\lim_{k\rightarrow\infty}\Psi(k)=\Psi$ , we have
\begin{equation}\label{5}
  \lim_{k\rightarrow\infty}(\Xi W)^{k}\Xi (W-I)=\lim_{k\rightarrow\infty}(\Xi W)^{k}(I-\Xi)=0.
\end{equation}
Let $\lim_{k\rightarrow\infty}(\Xi W)^{k}=Q=[q_{1}, q_{2}, \ldots, q_{n}]=[q_{ij}]_{n\times n}$ with $q_{i}\in\mathbb{R}^{n}$, from equation (\ref{5}) we obtain
\begin{equation*}
  \begin{cases}
  Q\Xi=Q\\
  QW=Q.
  \end{cases}
\end{equation*}
Because $Q\Xi=Q$, we have $q_{i}=0$ for any $i\in V_{f}\bigcup V_{p}$. It follows from $QW=Q$ that for any $i\in V, j\in V_{f}\bigcup V_{p}$, one has $\sum_{t=r_{1}+r_{2}+1}^{n}q_{it}w_{tj}=0$. Recalling that $Q$ and $W$ are both nonnegative, we have that for any $i\in V_{n}$, if there exists $j\in V_{f}\bigcup V_{p}$ such that $w_{ij}\neq0$, then $q_{li}=0$ for any $l\in V$, i.e., $q_{i}=0$. Furthermore, for aforementioned $i\in V_{n}, l\in V$ and $q_{li}=0$, if there exists $u\in V_{n}$ satisfying $w_{ui}\neq0$, then $q_{lu}w_{ui}=q_{li}=0$. Thus, $q_{u}=0$. This indicates that for any agent $i\in V_{n}$, as long as it receives information from any stubborn agents directly or indirectly, there holds $q_{i}=0$.

Let $\tilde{Q}=\lim_{k\rightarrow\infty}\sum_{t=0}^{k-1}(\Xi W)^{t}(I-\Xi)=[\tilde{q}_{1}, \tilde{q}_{2}, \ldots, \tilde{q}_{n}]$, then there holds $\Psi=Q+\tilde{Q}=[\psi_{1}, \psi_{2}, \ldots, \psi_{n}]=[q_{1}+\tilde{q}_{1}, q_{2}+\tilde{q}_{2}, \ldots, q_{n}+\tilde{q}_{n}]$. Since $\xi_{i}=1$ for any $i\in V_{n}$, we obtain $\tilde{q}_{i}=0$ for any $i\in V_{n}$, namely, $q_{i}=\psi_{i}$ for any $i\in V_{n}$. In conclusion, for any $i\in V_{n}$, if there exists $j\in V_{f}\bigcup V_{p}$ which has a directed path to agent $i$ in graph $\mathcal{G}(W)$, then $\psi_{i}=0$.
$\blacksquare$

In \cite{Javad14}, the authors study the equilibrium of system (\ref{FJ}). Under the assumption that the network topology is undirected and connected, they present the form of equilibrium taking advantage of the appropriately defined hitting probabilities of a random walk over the network. However, since their conditions for convergence require that $\rho(\Xi W)<1$, the equilibrium is only related to the initial opinions of stubborn agents. A similar assumption for directed graph is presented in \cite{Bullo14}. In Property \ref{P2}, we show that under the condition of Assumption \ref{A1}, if a non-stubborn agent can obtain information directly or indirectly from any stubborn agents, including fully stubborn and partially stubborn agents, then the ultimate opinion $x^{*}$ does not include the information of this non-stubborn agent's initial opinion. Otherwise, according to the proof of Theorem \ref{T1}, the ultimate opinion is also related to the initial opinions of non-stubborn agents which form ISCCs with no stubborn agents in it.

\begin{remark}
Property \ref{P1} and Property \ref{P2} also provide an interesting perspective on leadership in social networks \citep{Dyer09} and animal groups \citep{Couzin05}. In \cite{Wang07}, a generalized definition of ``leader'' and ``follower" is presented. A leader refers to the agent who determines the ultimate state of the system, and the rest are followers. This definition includes the notion of ``leader'' and ``follower'' in which a leader is the agent who sends information to others but never receives information. From Property \ref{P1}, $\Psi$ is a stochastic matrix, thus the element of $\psi_{i}$ represents the proportion of the initial opinion of agent $i$ in each agent's ultimate opinion. Property \ref{P2} suggests that if a non-stubborn agent wants to determine the group's ultimate opinion, i.e., to be a leader, it must ensure that it is not influenced by stubborn agents. If we regard initial opinions of stubborn agents as external information, then this result is consistent with the experimental results in \cite{Dyer09} and \cite{Couzin05} that the leaders emerging spontaneously in groups are usually the individuals who have more power in communication or external information. Furthermore, for any individual who has no external information, it can never be a leader unless it does not receive information of the individuals who have external information.
\end{remark}

\section{\bf Opinion formation of the F-J model over issue sequences}\label{s-FJs}

In this section, we study opinion formation of the F-J model over sequences consisting of interdependent issues. On one hand, most of the prior investigations on opinion dynamics are considered over a single issue. In practice, however, associations of individuals are constituted to deal with sequences of issues within particular domains, and successive issues in specific issue domain are always correlated. On the other hand, for individuals, related issues always arise repeatedly in practice. Empirical evidences show that a group of individuals who share a common set of experiences will hold highly similar beliefs or characteristics \citep{Joshi10}. Hence, it is significant to consider the formation of opinions over issue sequences.

Next, we shall briefly introduce the path-dependence theory. The path-dependence theory is originally studied in economic and institutional change \citep{North90, Page06}, and subsequently developed in cognitive psychology \citep{Egidi97, MacKay06}. In \cite{Egidi97}, path-dependence is observed in the behavioral experiments of human groups. The main point of path-dependence is ``history matters". That is, the decision one faces for any given circumstance are constrained by the decisions one has made in the past, even though past circumstance may no longer be relevant. This phenomenon is ubiquitous in technological innovation, institutional change and cognitive process of humans, etc. In the cognitive and decision-making process of humans, people always have cognitive inertia with which they are unwilling to change their beliefs or decisions over interdependent issues. For example, one may wonder that why the U.S. standard railroad gauge is $4$ feet $8.5$ inches, which is an exceedingly odd number. A well known explanation for this question is that it is resulted from path-dependence \citep{North90}. Firstly, the U.S. railroads were built by English expatriates and that is the guage they built them in England. Further, the first rail lines of England were built by the same people who built the pre-railroad tramways, where that gauge was used. Moreover, the tramways were built by the same people who built wagons using the same jigs and tools. In fact, that gauge is the wheel spacing of the wagon. In this example, people's decisions over the interdependent issue sequence ``The gauge of the U.S. railroad'', ``The gauge of the English railroad'', ``The gauge of the tramway'' and ``The wheel spacing of the wagon'' remain unaltered. This phenomenon is named cognitive freezing \citep{MacKay06}. Since in the F-J model, agents' initial opinions can be regarded as their internal opinions, we will establish a connection between agents' initial opinions for interdependent issues based on the path-dependence theory.
\begin{figure}
 \centering
  \includegraphics[width=8cm]{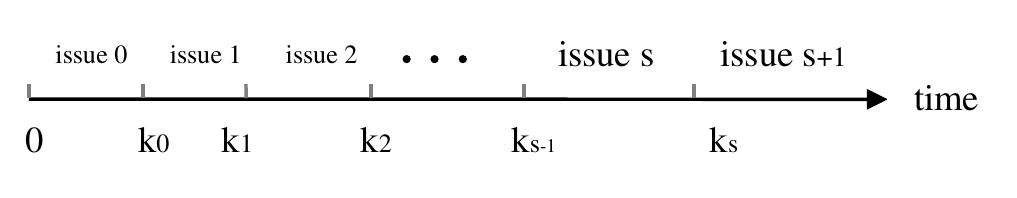}\\
  \caption{ An issue sequence.}\label{figi}
\end{figure}

Let $s=0, 1, 2,\ldots$ denote a sequence of interdependent or repeatedly arising issues over which path-dependence works\footnote{Note that the issue sequence is not necessarily consecutive over time, some other uncorrelated issues may interpolate in it. For example, Let $S_{1}=S_{1}^{0}, S_{1}^{1}, S_{1}^{2},\ldots$ denote a issue sequence which consists of interdependent issues, and $S_{2}$ denote another. The issues in $S_{1}$ and $S_{2}$ may be irrelevant and arise alternately. Without loss of generality, we focus on each path-dependent subsequence in this paper.}, $k_{s}$ be a constant denoting the time at which issue $s$ arises, $x(s, k)$ denote agents' opinions for issue $s$ at time $k$. Suppose that issue $0$ arises at time $0$, agents stop dealing with issue $s-1$ and start to deal with issue $s$ at time $k_{s}$ (Fig. \ref{figi}). Note that different from the assumptions in \cite{Bullo14} and \cite{Bullo15} which require the time interval $k_{s+1}-k_{s}$ tends to infinity, here $k_{s+1}-k_{s}$ can be any positive integer. For a given issue $s$, system (\ref{FJ}) is modified as:
\begin{equation}\label{FJs}
  x(s, k+1)=\Xi Wx(s, k)+(I-\Xi)x(s, k_{s}),
\end{equation}
where $x(s, k_{s})$ represents initial opinions, i.e., internal opinions, of agents for issue $s$, and $k\in[k_{s},k_{s+1}]$.

Let $\zeta_{i}$ denote the factor of cognitive inertia of agent $i$ for two successive issues in an path-dependent issue sequence. Note that the basic assumption of the F-J model is that agents are anchored to their initial opinions over a given issue, this is coincident with the path-dependence theory which emphasizes that individual has the cognitive inertia over a sequence of interdependent issues. Thus, we assume that for each agent $i$, its factor of cognitive inertia $\zeta_{i}$ equals to its stubborn factor, i.e., $\zeta_{i}=1-\xi_{i}$. Then, basing on the game theory formulation of the F-J model in \cite{Bindel11} and the self-reinforcing process of path-dependence \citep{North90}, we assume that in forming its initial opinion for issue $s+1$, agent $i$ attempts to myopically minimize the following cost function:
\begin{equation}\label{FJs1.2}
C_{i}(x)=\zeta_{i}(x_{i}-x_{i}(s,k_{s}))^2+(1-\zeta_{i})\sum_{j=1}^{n}w_{ij}(x_{i}-x_{j})^2,
\end{equation}
where the first term of $C_{i}(x)$ depicts the cognitive inertia of agent $i$ to change its initial opinions (i.e., internal opinions) for interdependent issues, and the second term represents the willingness of agent $i$ to escape being unsocial. As a consequence, the initial opinion of agent $i$ for issue $s+1$ satisfies
\begin{equation}\label{FJs1.1}
x_{i}(s+1, k_{s+1})=\zeta_{i}x_{i}(s,k_{s})+(1-\zeta_{i})\sum_{j=1}^{n}w_{ij}x_{j}(s+1,k_{s+1}),
\end{equation}
i.e.,
\begin{equation}\label{FJs1}
x(s+1, k_{s+1})=\Lambda x(s,k_{s})+(I-\Lambda)Wx(s+1,k_{s+1}),
\end{equation}
where $\Lambda=diag(\zeta_{1},\zeta_{2},\ldots,\zeta_{n})^{T}$.

The hypothesis in equation (\ref{FJs1}) can be naturally observed from reality. In practice, an individual always has cognitive inertia over sequences of path-dependent issues, especially when an issue arises repeatedly. In extreme cases, cognitive inertia may reinforce into cognitive freezing with which individuals keep their decisions unchangeable, even they are exposed to more efficient options \citep{Egidi97, MacKay06}. Next we propose an assumption about $\mathcal{G}(W)$ which ensures that $I-\Xi W$ is nonsingular.

\begin{assumption}\label{A2}
For each ISCC of $\mathcal{G}(W)$, there exists at least one stubborn agent in it.
\end{assumption}
\begin{lemma}\label{L2}
$I-\Xi W$ is nonsingular if and only if Assumption \ref{A2} holds.
\end{lemma}
{\bf Proof.} {\it Sufficiency}. Since Assumption \ref{A2} holds, the proof of Theorem \ref{T1} shows that $\rho(\Xi W)=\rho(\tilde{\Xi}\tilde{W})<1$, which implies that $I-\Xi W$ is nonsingular.

{\it Necessity}. If Assumption \ref{A2} does not hold, there exists at least one eigenvalue of $\Xi W$ equals to $1$. Namely, $I-\Xi W$ is singular. $\blacksquare$

From equation (\ref{FJs1}) we have
\begin{equation*}
(I-W+\Lambda W)x(s+1, k_{s+1})=\Lambda x(s, k_{s}).
\end{equation*}
If Assumption \ref{A2} holds, then $I-\Xi W$ is nonsingular. Since $\Lambda=I-\Xi$, we obtain
\begin{equation}\label{FJs1.3}
x(s+1, k_{s+1})=(I-\Xi W)^{-1}(I-\Xi)x(s, k_{s}).
\end{equation}
Since $\Xi W+I-\Xi$ is a stochastic matrix, we have that for any nontrivial $\Xi$ and $W$, $x(s,k_{s+1})=\chi \mathbf{1}_{n}$ if and only if $x(s,k_{s})=\chi \mathbf{1}_{n}$, where $\chi\in\mathbb{R}$ is a constant. Thus, we focus on consensus of agents' initial opinions for each issue in the sequel. Now, we shall propose a definition of opinion consensus for system (\ref{FJs}), (\ref{FJs1}). 

\begin{definition}\label{D2}
System (\ref{FJs}), (\ref{FJs1}) is said to achieve consensus if and only if for any initial opinion $x(0,0)$, there exists a constant $\chi\in\mathbb{R}$, such that
\begin{equation}\label{d2}
  \lim_{s\rightarrow\infty}x(s,k_{s})=\chi \mathbf{1}_{n}.
\end{equation}
\end{definition}

\begin{remark}\label{R4}
Suppose that Assumption \ref{A2} holds. If $k_{s+1}-k_{s}$ is large enough for each $s$, such that
\begin{equation*}
x(s,k_{s+1})=x(s,\infty),
\end{equation*}
then, according to Lemma \ref{L2} and equation (\ref{FJs}), we obtain
\begin{equation*}
x(s,\infty)=(I-\Xi W)^{-1}(I-\Xi)x(s,k_{s}).
\end{equation*}
Thus,
\begin{equation}\label{FJs2}
 x(s+1,k_{s+1})=x(s,\infty),
\end{equation}
which implies that agents' initial opinions for issue $s+1$ equal to their limiting opinions for issue $s$, namely, cognitive freezing happens. This is consistent with our discussion above that cognitive inertia can be reinforced into cognitive freezing in extreme cases. For example, as more people bought QWERTY keyboard, buying other keyboards with different key configuration requires learning to type anew. Thus, people's opinions for the sequence of issues related to keyboard will be locked in QWERTY keyboard, even though there are many other more efficient types of keyboard \citep{Page06}. Other examples of cognitive freezing are also reported, such as the fall of the Dundee jute industry \citep{MacKay06}, the persistence of the narrow-gauge rail \citep{North90}, etc. In \cite{Page06}, the author attributes cognitive freezing to the externalities of other people. Note that in equation (\ref{FJs1.2}), we assume that each agent's initial opinion for the next issue is influenced by other agents, i.e., the interpersonal influence creates externalities under Assumption \ref{A2}, which lead to cognitive freezing in equation (\ref{FJs2}).
\end{remark}

Let $\Psi=[\psi_{ij}]_{n\times n}=(I-\Xi W)^{-1}(I-\Xi)$. The next lemma reveals the connections between $\mathcal{G}(\Psi)$ and $\mathcal{G}(W)$.
\begin{lemma}\label{L3}
Under Assumption \ref{A2}, the following statements hold:\\
(i) $\psi_{ii}>0$ for any $i\in V_{f}\bigcup V_{p}$.\\
(ii) For any $i\in V_{p}\bigcup V_{n}, j\in V_{f}\bigcup V_{p}, i\neq j$, $\psi_{ij}>0$ if and only if there exists a directed path $(j, j_{1}), (j_{1}, j_{2}),\ldots, (j_{m}, i)$ from vertex $j$ to vertex $i$ in $\mathcal{G}(W)$, where $j_{1}, j_{2},\ldots,j_{m}\in V_{p}\bigcup V_{n}$.
\end{lemma}
{\bf Proof.} Let $\hat{\Psi}=[\hat{\psi}_{ij}]_{n\times n}=(I-\Xi W)^{-1}$.\\
(i) Since $(I-\Xi W)\hat{\Psi}=I$, we have
\begin{equation}\label{FJs3}
  \hat{\psi}_{ii}(1-\xi_{i}w_{ii})=1+\xi_{i}\sum_{\substack{t=1 \\ t\neq i}}^{n}w_{it}\hat{\psi}_{ti}, \ \ \ i \in V.
\end{equation}

Assumption \ref{A2} implies that if $w_{ii}=1$, $\xi_{i}<1$. Thus, $\xi_{i}w_{ii}<1$ for any $i\in V$. From (\ref{FJs3}) we obtain
\begin{equation*}
 \hat{\psi}_{ii}=\frac{1+\xi_{i}\sum\limits_{\substack{t=1 \\ t\neq i}}^{n}w_{it}\hat{\psi}_{ti}}{1-\xi_{i}w_{ii}}\geq \frac{1}{1-\xi_{i}w_{ii}}>0.
\end{equation*}
Since $1-\xi_{i}>0$ for any $i\in V_{f}\bigcup V_{p}$, we have $\psi_{ii}=\hat{\psi}_{ii}(1-\xi_{i})>0$ for any $i\in V_{f}\bigcup V_{p}$.

(ii) Since Assumption \ref{A2} holds, one has $(I-\Xi W)^{-1}=\sum\limits_{t=0}^{\infty}(\Xi W)^{t}$. Let $(\Xi W)^{t}=[\hat{\psi}^{t}_{ij}]_{n\times n}$. Note that $\hat{\psi}^{t}_{ij}>0$ represents that there exist paths consisting of $t$ edges (maybe appear repeatedly or contain self-loops ) in $\mathcal{G}(\Xi W)$ start from vertex $j$ to vertex $i$. Moreover, if we take no account of the weights, $\mathcal{G}(\Xi W)$ and $\mathcal{G}(W)$ have no difference except the edges to fully stubborn agents. Thus, if there exists a directed path $(j, j_{1}), (j_{1}, j_{2}),\ldots, (j_{m}, i)$ from vertex $j$ to vertex $i$ in $\mathcal{G}(W)$, and $j_{1}, j_{2},\ldots,j_{m}\in V_{p}\bigcup V_{n}$, we have $\hat{\psi}^{m}_{ij}>0$. Therefore, $\hat{\psi}_{ij}>0$. Otherwise, we have $\hat{\psi}^{t}_{ij}=0$ for any $t\in\mathbb{N}$. Thus, $\hat{\psi}_{ij}=0$. Since $\xi_{j}<1$ for $j\in V_{f}\bigcup V_{p}$, $\psi_{ij}=\hat{\psi}_{ij}(1-\xi_{j})$, we have that $\psi_{ij}>0$ for any $i\in V_{p}\bigcup V_{n}, j\in V_{f}\bigcup V_{p}, i\neq j$ if and only if there exists a directed path $(j, j_{1}), (j_{1}, j_{2}),\ldots, (j_{m}, i)$ from vertex $j$ to vertex $i$ in $\mathcal{G}(W)$, where $j_{1}, j_{2},\ldots,j_{m}\in V_{p}\bigcup V_{n}$. $\blacksquare$

Lemma \ref{L3} connects $\mathcal{G}(\Psi)$ with $\mathcal{G}(W)$, such that we can address property of $\mathcal{G}(\Psi)$ by analyzing $\mathcal{G}(W)$. From the proof of Lemma \ref{L3}, we have
\begin{equation*}
 \Psi=\begin{bmatrix}
  I_{r_{1}\times r_{1}} & 0 & 0\\
  \Psi_{pf} & \Psi_{pp} & 0\\
  \Psi_{nf} & \Psi_{np} & 0
  \end{bmatrix},
\end{equation*}
where $\Psi_{pp}\in\mathbb{R}^{r_{2}\times r_{2}}$.
\begin{property}\label{P3}
Any root of subgraph $\mathcal{G}(\Psi_{pp})$ is a center vertex which has directed edges to any other vertices of $\mathcal{G}(\Psi_{pp})$.
\end{property}
{\bf Proof.} Suppose that $i$ is a root of $\mathcal{G}(\Psi_{pp})$. Let $(i,j)$ and $(j,l)$ be edges of $\mathcal{G}(\Psi_{pp})$ for $j, l\in V_{p}$. By Lemma \ref{L3}, there exist directed paths consisting of vertices in $V_{p}\bigcup V_{n}$ from $i$ to $j$ and $j$ to $l$ in $\mathcal{G}(W)$, respectively. Thus, there exists at least a directed path consisting of edges between vertices in $V_{p}\bigcup V_{n}$ from $i$ to $l$ in $\mathcal{G}(W)$. From Lemma \ref{L3} we have $\psi_{li}>0$, namely, there exists a directed edge from $i$ to $l$ in $\mathcal{G}(\Psi_{pp})$. Following the same lines as above, one has that any root of subgraph $\mathcal{G}(\Psi_{pp})$ is a center vertex which has directed edges to any other vertices of $\mathcal{G}(\Psi_{pp})$. $\blacksquare$

Lemma \ref{L3} and Property \ref{P3} show that if there exists a directed path from a stubborn agent $i$ to a partially stubborn or non-stubborn agent $j$ in $\mathcal{G}(W)$, then $(i,j)$ is an edge of $\mathcal{G}(\Psi)$. Moreover, if $\mathcal{G}(\Psi_{pp})$ contains a spanning tree, then it also contains a star subgraph. In other words, the connectivity of the social network is enhanced over issue sequences. The block form of $\Psi$ shows that fully stubborn agents never change their opinions in the process of opinion formation. Thus, we consider the scenario in which there are no fully stubborn agents in the social network.
\begin{theorem}\label{T2}
Suppose that $V_{f}=\varnothing$ and Assumption \ref{A2} holds. System (\ref{FJs}), (\ref{FJs1}) achieves opinion consensus if and only if there exists a partially stubborn agent which has directed paths to any other partially stubborn agents in $\mathcal{G}(W)$.
\end{theorem}
{\bf Proof.} {\it Sufficiency}. Since $V_{f}=\varnothing$, i.e., $r_{1}=0$, we have
\begin{equation*}
   W=\begin{bmatrix}
  W_{pp} & W_{pn}\\
  W_{np} & W_{nn}
  \end{bmatrix}, \Psi=\begin{bmatrix}
   \Psi_{pp} & 0\\
   \Psi_{np} & 0
  \end{bmatrix},
\end{equation*}
where $W_{pp}\in\mathbb{R}^{r_{2}\times r_{2}}, W_{nn}\in\mathbb{R}^{n-r_{2}\times n-r_{2}}$, $\Psi_{pp}\in\mathbb{R}^{r_{2}\times r_{2}}$. From Property \ref{P1} we have that $\Psi_{pp}$ and $\Psi_{np}$ are both stochastic matrices. Lemma \ref{L3}. (i) indicates that $\Psi_{pp}$ has positive diagonal elements. Moreover, since there exists a partially stubborn agent which has directed paths consisting of edges between vertices in $V_{p}\bigcup V_{n}$ to any other partially stubborn agents in $\mathcal{G}(W)$, Lemma \ref{L3}. (ii) implies that $\mathcal{G}(\Psi_{pp})$ has a spanning tree. By Corollary 3.5 and Lemma 3.7 in \cite{ren05}, we have that $\Psi_{pp}$ is stochastic, indecomposable and aperiodic (SIA). That is, there exists a nonnegative vector $\nu\in\mathbb{R}^{r_{2}}$ such that $\lim\limits_{s\rightarrow\infty}\Psi_{pp}^{s}=\mathbf{1}_{r_{2}}\nu^{T}$, where $\nu$ satisfies $\Psi_{pp}^{T}\nu=\nu$ and $\mathbf{1}^{T}\nu=1$. Note that
\begin{equation*}
  \lim_{s\rightarrow\infty}\Psi^{s}=\lim_{s\rightarrow\infty}\begin{bmatrix}
   \Psi_{pp}^{s} & 0\\
   \Psi_{np}\Psi_{pp}^{s-1} & 0
  \end{bmatrix}=\begin{bmatrix}
    \mathbf{1}_{r_{2}}\nu^{T}& 0\\
   \Psi_{np}\mathbf{1}_{r_{2}}\nu^{T} & 0
  \end{bmatrix}=\begin{bmatrix}
    \mathbf{1}_{r_{2}}\nu^{T}& 0\\
    \mathbf{1}_{n-r_{2}}\nu^{T} & 0
  \end{bmatrix},
\end{equation*}
which indicates that $\Psi_{pp}^{s}$ converges to a matrix of rank one as $s\rightarrow\infty$. Since Assumption \ref{A2} holds, we have $\lim_{s\rightarrow\infty}x(s,k_{s})=\Psi^{s}x(0,0)=[\nu^{T} \ \mathbf{0}_{n-r_{2}}^{T}]x(0,0)\mathbf{1}_{n}$, namely, system (\ref{FJs}), (\ref{FJs1}) achieves opinion consensus as $s\rightarrow\infty$.

{\it Necessity}. If there exists no partially stubborn agent which has directed paths consisting of vertices in $V_{p}\bigcup V_{n}$ to any other partially stubborn agents in $\mathcal{G}(W)$, Lemma \ref{L3}. (ii) suggests that $\mathcal{G}(\Psi_{pp})$ does not contain any spanning trees. This implies that the eigenvalue $1$ of $\Psi_{pp}$ has algebraic multiplicity larger than one, that is, $\Psi$ has more than one eigenvalue equal to $1$. In conclusion, opinion consensus cannot be achieved. $\blacksquare$

As is often pointed out, the F-J model does not generally achieve opinion consensus over a single issue even if the network topology is completely connected \citep{Javad14, Frasca15}. However, Theorem \ref{T2} shows that in the absence of fully stubborn agents, as long as there exists at least one partially stubborn agent whose opinion can scatter to other partially stubborn agents through partially stubborn agents or non-stubborn agents, opinion consensus can be achieved over a sequence of issues. The block form of $\Psi$ indicates that when there exists more than one fully stubborn agent in the network, consensus can not be achieved for arbitrary initial opinion unless the influence of fully stubborn agents to others is identical \citep{Friedkin15}. Since the condition for achieving opinion consensus in Theorem \ref{T2} is sufficient and necessary, it also gives the condition for opinions of agents to form clusters.

\begin{corollary}\label{C1}
Suppose that $V_{f}=\varnothing$ and Assumption \ref{A2} holds. System (\ref{FJs}), (\ref{FJs1}) forms opinion clusters if and only if there exists more than one ISCC in $\mathcal{G}(W)$.
\end{corollary}

\section{\bf Opinion formation of the F-J model over issue sequences with bounded confidence}\label{s-FJh}
As discussed in Section \ref{s-FJs}, the cognitive inertia of agent can be reinforced into cognitive freezing under Assumption \ref{A2}, i.e., each agent's initial opinion for issue $s+1$ equals to its limiting opinion for issue $s$ if its opinion for $s$ converges. It is mainly because of the externalities created by agents' interpersonal influence under which agents treat opinions of others with no difference. However, in practice, people usually assimilate information in a selective way: they tend to give considerable weight to the information supporting their initial opinions, and dismiss the undermining information meanwhile, which is named echo-chamber effect \citep{Sunstein12, Dandekar13}. Similar observations in animal groups are reported in \cite{Couzin05}. In this section, we consider a more general scenario that each agent maintains a confidence bound in forming its initial opinions.

For simplicity, we employ the assumption used in \cite{Bullo14} and \cite{Bullo15} that the scale of time $k$ increases faster than the scale of issue $s$ in this section. In other words, the convergence of opinion $x(s,k)$ can be achieved with $k$ increasing before the next issue $s+1$ arising, i.e., $k_{s}=\infty$ for issue $s-1$ and $k_{s}=0$ for issue $s$. From Remark \ref{R4}, we have $x(s+1, 0)=x(s,\infty)$ under Assumption \ref{A2}, which means that agents take their ultimate opinions for the last issue as their initial opinions for the next issue. Inspired by the echo-chamber effect, we suppose that agent $i$ forms its initial opinion $x_{i}(s+1,0)$ by taking a weighted average of other agents' ultimate opinions $x_{j}(s,\infty), j\in V$ which satisfy $|x_{i}(s,\infty)-x_{j}(s,\infty)|<d$, where $d\in\mathbb{R}$ is the given confidence bound.

Let $\mathcal{N}_{i}(s)=\{j\in V:|x_{i}(s,\infty)-x_{j}(s,\infty)|<d\}$ denote the set of agents whose ultimate opinions for issue $s$ are within the confidence interval of agent $i$, $\mathcal{N}_{i}^{\Psi}=\{j\in V: \psi_{ij}>0\}$ denote the set of neighbors of agent $i$ in $\mathcal{G}(\Psi)$. Here we employ a symmetric modification of the H-K model \citep{J14, Yang14}:
\begin{equation}\label{FJh}
 x_{i}(s+1,0)=x_{i}(s,\infty)+h\sum_{j\in\mathcal{N}_{i}(s)}(x_{j}(s,\infty)-x_{i}(s,\infty)),
\end{equation}
where $h\in\mathbb{R}$ satisfies $h>0$ and $1-(n-1)h>0$. Note that if the differences of agents' initial opinions are large enough such that $\mathcal{N}_{i}(s)=\{i\}$ for any $i\in V$ and $s\in\mathbb{N}$, equation (\ref{FJh}) will collapse to equation (\ref{FJs1.3}). Combining equation (\ref{FJh}) and Lemma \ref{L2}, we have
\begin{equation*}
  x(s,\infty)=\Psi x(s,0),
\end{equation*}
and
\begin{equation*}
  x(s+1,0)=H(s)x(s,\infty),
\end{equation*}
where $\Psi=(I-\Xi W)^{-1}(I-\Xi)$, $H(s)=[h_{ij}(s)]_{n\times n}$, and
\begin{equation}\label{FJh1}
  \begin{cases}
  h_{ii}(s)=1-h(|\mathcal{N}_{i}(s)|-1) & i\in V,\\
  h_{ij}(s)=h & j\neq i, j\in \mathcal{N}_{i}(s),\\
  h_{ij}(s)=0 & otherwise.
  \end{cases}
\end{equation}
For simplicity, we omit the time scale $k$ in the sequel. Let $\Phi(s)=[\varphi_{ij}(s)]_{n\times n}=H(s)\Psi$, then we have $x(s+1)=\Phi(s)x(s)$. According to the partition of $V$, we have
\begin{equation*}
  \Phi(s)=\begin{bmatrix}
  \Phi_{1}(s) & 0\\
  \Phi_{2}(s) & 0
  \end{bmatrix},
\end{equation*}
where $\Phi_{1}(s)\in\mathbb{R}^{(r_{1}+r_{2})\times (r_{1}+r_{2})}$. Rewrite $H(s)$ as a same block form of $W$:
\begin{equation*}
  H(s)=\begin{bmatrix}
  H_{ff}(s) & H_{fp}(s) & H_{fn}(s)\\
  H_{pf}(s) & H_{pp}(s) & H_{pn}(s)\\
  H_{nf}(s) & H_{np}(s) & H_{nn}(s)
  \end{bmatrix},
\end{equation*}
where $H_{ff}(s)\in\mathbb{R}^{r_{1}\times r_{1}}$, $H_{pp}(s)\in\mathbb{R}^{r_{2}\times r_{2}}$ and $H_{nn}(s)\in\mathbb{R}^{(n-r_{1}-r_{2})\times (n-r_{1}-r_{2})}$. Let
\begin{equation*}
  \Phi^{\prime}(s)=\begin{bmatrix}
  H_{ff}(s) & H_{fp}(s)\\
  W_{pf} & W_{pp}
  \end{bmatrix}_{(r_{1}+r_{2})\times (r_{1}+r_{2})}.
\end{equation*}
\begin{lemma}\label{L4}
For any $s\in\mathbb{N}$, if $\mathcal{G}(\Phi^{\prime}(s))$ has a spanning tree, then $\Phi_{1}(s)$ is SIA.
\end{lemma}
{\bf Proof.} From equation (\ref{FJh1}), $H(s)$ is a nonnegative and stochastic matrix with positive diagonal elements. By Lemma \ref{L3}, we have $\psi_{ii}>0$ for any $i\in V_{f}\bigcup V_{p}$. Thus, $\Phi(s)$ and $\Phi_{1}(s)$ are both stochastic matrices and $\Phi_{1}(s)$ has positive diagonal elements. Since $\Phi(s)=H(s)\Psi$, we have
\begin{equation}\label{FJh2}
  \begin{cases}
  \varphi_{ij}(s)=h_{ij}(s)+\sum\limits_{t=r_{1}+1}^{n}h_{it}(s)\psi_{tj} & i\in V_{f}\bigcup V_{p}, j\in V_{f},\\
  \varphi_{ij}(s)=\sum\limits_{t=r_{1}+1}^{n}h_{it}(s)\psi_{tj} & i\in V_{f}\bigcup V_{p}, j\in V_{p}.
  \end{cases}
\end{equation}
Hence, for any $i\in V_{f}, j\in V_{f}\bigcup V_{p}$, $h_{ij}(s)>0$ indicates $\varphi_{ij}(s)>0$. Moreover, Lemma \ref{L3}. (ii) shows that for $i\in V_{p}, j\in V_{f}\bigcup V_{p}$, $w_{ij}>0$ means $\psi_{ij}>0$. Thus, for any $i\in V_{p}, j\in V_{f}\bigcup V_{p}$, $w_{ij}>0$ implies $\varphi_{ij}>0$. Because $\mathcal{G}(\Phi^{\prime}(s))$ has a spanning tree, we have $\mathcal{G}(\Phi_{1}(s))$ contains a spanning tree. Following from Corollary 3.5 and Lemma 3.7 in \cite{ren05}, $\Phi_{1}(s)$ is SIA. $\blacksquare$

Lemma \ref{L4} implies that as long as the connectivity of $\mathcal{G}(\Phi^{\prime}(s))$ can be preserved, $\Phi_{1}(s)$ is SIA. This is on account of enhanced connectivity of the interpersonal influence network over issue sequences, which is shown in Property \ref{P3}. Next, we give the condition for $H(0)$ to preserve connectivity.

Let $y(s)=\Psi x(s)$, then $x(s+1)=H(s)y(s)$. Denote the set of common neighbors of agents $i$ and $j$ for issue $s$ by
\begin{equation*}
 \mathcal{N}_{ij}(s)=\{l\in V:|y_{i}(s)-y_{l}(s)|<d, |y_{j}(s)-y_{l}(s)|<d \}.
\end{equation*}
From equation (\ref{FJh}), we have that $\mathcal{G}(H(s))$ is undirected. The following assumption proposes the conditions for connectivity preservation of $\mathcal{G}(H(s))$.
\begin{assumption}\label{A3}
(i) For any $i\in V_{f}, j\in(V_{p}\bigcup V_{n})\bigcap\mathcal{N}_{i}(0)$ and $l\in\mathcal{N}_{j}^{\Psi}$, $l\in\mathcal{N}_{i}(0)$. For any $i\in V_{p}\bigcup V_{n}, j\in(V_{p}\bigcup V_{n})\bigcap\mathcal{N}_{i}(0)$ and $l\in\mathcal{N}_{j}^{\Psi}$, $l\in\mathcal{N}_{j}(0)\bigcap\mathcal{N}_{i}(0)$.\\
(ii) For any $i\in V_{f}, j\in V_{f}\bigcap\mathcal{N}_{i}(0)$, $|\mathcal{N}_{ij}(0)|>\frac{n}{2}$. For any $i\in V_{p}\bigcup V_{n}, j\in V\bigcap\mathcal{N}_{i}(0)$, $|\mathcal{N}_{ij}(0)|>\frac{n}{2}+\frac{1}{4h}$.
\end{assumption}

As mentioned above, only if the difference between two agents' opinions is small enough, agents will take account of other agents' opinions. In Assumption \ref{A3}, we give some conditions about the distribution of agents' ultimate opinions for the first issue to ensure that $\mathcal{G}(H(0))$ is adequately connected and thus the connectivity of $\mathcal{G}(H(s))$ can be preserved for any $s\in\mathbb{N}$. Note that given a social network, the interpersonal influence network $\mathcal{G}(W)$ and agents' stubborn extent $\Xi$ are determined, and $y(0)=(I-\Xi W)^{-1}(I-\Xi)x(0,0)$. Therefore, the conditions on $y(0)$ in Assumption \ref{A3} are actually conditions on $x(0,0)$. From equation (\ref{FJh}), we have $h<\frac{1}{n-1}$. Moreover, since $|\mathcal{N}_{ij}(0)|<n$, from Assumption \ref{A3} we have $h>\frac{1}{2n}$. Thus, $\frac{1}{2n}<h<\frac{1}{n-1}$.

\begin{lemma}\label{L5}
Under Assumption \ref{A3}, no edge in $\mathcal{G}(H(0))$ will be lost in $\mathcal{G}(H(s))$ for any $s\in\mathbb{N}$.
\end{lemma}
{\bf Proof.} Firstly, we shall show that if Assumption \ref{A3} holds for issue $s$, then it holds for issue $s+1$. Suppose that for any $|y_{i}(s)-y_{j}(s)|<d$, $i, j\in V$, Assumption \ref{A3} holds for issue $s$. For $i, j\in V_{f}$, we have
\begin{equation*}
\begin{split}
  &|y_{i}(s+1)-y_{j}(s+1)|\\
 =&|x_{i}(s+1)-x_{j}(s+1)|\\
 =&\bigg|y_{i}(s)+h\sum_{l\in\mathcal{N}_{i}(s)}(y_{l}(s)-y_{i}(s))-y_{j}(s)-h\sum_{l\in\mathcal{N}_{j}(s)}(y_{l}(s)-y_{j}(s))\bigg|\\
 =&\bigg|y_{i}(s)-y_{j}(s)+h\sum_{l\in\mathcal{N}_{ij}(s)}(y_{j}(s)-y_{i}(s))+\\
  &h\sum_{l\in\mathcal{N}_{i}(s)\backslash\mathcal{N}_{ij}(s)}(y_{l}(s)-y_{i}(s))-h\sum_{l\in\mathcal{N}_{j}(s)\backslash\mathcal{N}_{ij}(s)}(y_{l}(s)-y_{j}(s))\bigg|\\
  \leq&(1-h|\mathcal{N}_{ij}(s)|)|y_{i}(s)-y_{j}(s)|+h\sum_{l\in\mathcal{N}_{i}(s)\backslash\mathcal{N}_{ij}(s)}|(y_{l}(s)-y_{i}(s))|\\
  &+h\sum_{l\in\mathcal{N}_{j}(s)\backslash\mathcal{N}_{ij}(s)}|(y_{l}(s)-y_{j}(s))|\\
  <&(1-h|\mathcal{N}_{ij}(s)|)d+h(|\mathcal{N}_{i}(s)|+|\mathcal{N}_{j}(s)|-2|\mathcal{N}_{ij}(s)|)d\\
  \leq&(1-h|\mathcal{N}_{ij}(s)|)d+h(n-|\mathcal{N}_{ij}(s)|)d\\
  \leq&(1+hn-2h|\mathcal{N}_{ij}(s)|)d\\
  <&d.
\end{split}
\end{equation*}
For $i\in V_{f}, j\in V_{p}\bigcup V_{n}$, we have
\begin{equation*}
\begin{split}
   &|y_{i}(s+1)-y_{j}(s+1)|\\
  =&\bigg|x_{i}(s+1)-\sum_{l\in\mathcal{N}_{j}^{\Psi}}\psi_{jl}x_{l}(s+1)\bigg|\\
  =&\bigg|\sum_{l\in\mathcal{N}_{j}^{\Psi}}\psi_{jl}(x_{i}(s+1)-x_{l}(s+1))\bigg|.
\end{split}
\end{equation*}
Noting that for any $l\in\mathcal{N}_{j}^{\Psi}$, one has $l\in\mathcal{N}_{i}(s)$ and $|\mathcal{N}_{il}(0)|>\frac{n}{2}$. Following the same lines as above, we obtain
\begin{equation*}
\begin{split}
  |y_{i}(s+1)-y_{j}(s+1)\mid<\sum_{l\in\mathcal{N}_{j}^{\Psi}}\psi_{jl}d=d.
\end{split}
\end{equation*}
For $i, j\in(V_{p}\bigcup V_{n})$, we get
\begin{equation*}
\begin{split}
  |y_{i}(s+1)-y_{j}(s+1)|=\bigg|\sum_{l\in\mathcal{N}_{i}^{\Psi}}\psi_{il}x_{l}(s+1)-\sum_{l\in\mathcal{N}_{j}^{\Psi}}\psi_{jl}x_{l}(s+1)\bigg|.
\end{split}
\end{equation*}
Since for any $l\in\mathcal{N}_{j}^{\Psi}$, Assumption \ref{A3} ensures $l\in\mathcal{N}_{j}(s)\bigcap\mathcal{N}_{i}(s)$ and $|\mathcal{N}_{il}(0)|>\frac{n}{2}+\frac{1}{4h}$, we obtain
\begin{equation*}
\begin{split}
    &\bigg|\sum_{l\in\mathcal{N}_{i}^{\Psi}}\psi_{il}x_{l}(s+1)-\sum_{l\in\mathcal{N}_{j}^{\Psi}}\psi_{jl}x_{l}(s+1)\bigg|\\
   =&\bigg|\sum_{l\in\mathcal{N}_{i}^{\Psi}}\psi_{il}(x_{l}(s+1)-x_{i}(s+1))\\
    &+\sum_{l\in\mathcal{N}_{j}^{\Psi}}\psi_{jl}(x_{i}(s+1)-x_{l}(s+1))\bigg|\\
\leq&\bigg|\sum_{l\in\mathcal{N}_{i}^{\Psi}}\psi_{il}(x_{l}(s+1)-x_{i}(s+1))\bigg|\\
    &+\bigg|\sum_{l\in\mathcal{N}_{j}^{\Psi}}\psi_{jl}(x_{i}(s+1)-x_{l}(s+1))\bigg|\\
   <&\sum_{l\in\mathcal{N}_{i}^{\Psi}}\psi_{il}\frac{d}{2}+\sum_{l\in\mathcal{N}_{j}^{\Psi}}\psi_{jl}\frac{d}{2}=d.
\end{split}
\end{equation*}
That is to say, if Assumption \ref{A3} holds for issue $s$, then no edge contained in $\mathcal{G}(H(s))$ will be lost, which implies that Assumption \ref{A3} holds for issue $s+1$. Thus, no edge in $\mathcal{G}(H(0))$ will be lost in $\mathcal{G}(H(s))$ for any $s\in\mathbb{N}$ under Assumption \ref{A3}. $\blacksquare$

The connectivity preservation is a crucial and challenging problem in research for the consensus problem of the H-K model since the evolution of agents' opinions is only associated with the initial states and their distribution. In \cite{J14}, a potential function is presented to measure the total difference of agents' initial opinions such that the connectivity is preserved by restricting the distribution of agents' initial opinions. Following a different line, in Lemma \ref{L5} we ensure that the links existing in $\mathcal{G}(H(0))$ will never be lost as $s$ increases by proposing a lower bound for the number of agents' common neighbors in $\mathcal{G}(H(0))$. In addition, different from the modified H-K model considered in \cite{Yang14}, in system (\ref{FJs}), (\ref{FJh}), the evolution of opinions over time scale increases the complexity of our analysis.

\begin{lemma}\label{L4.1}
(Lemma 3.1, \cite{ren05}) Let $P_{1}, P_{2},...$, $P_{m}$ $\in\mathbb{R}^{n\times n}$ be $m$ nonnegative matrices with positive diagonal elements, then
\begin{equation*}
  P_{1} P_{2}...P_{m}\geq\eta(P_{1}+P_{2}+...+P_{m}),
\end{equation*}
where $\eta$ can be specified from matrices $P_{i}, i=1,2,...,m$.
\end{lemma}

By the proof of Lemma \ref{L4} and Lemma \ref{L5}, that the subgraph of $\mathcal{G}(H(0))$ corresponding to stubborn agents is connected also implies that $\Phi_{1}(s)$ is SIA. In the next theorem, we employ a more feasible condition which reduces the requirements for the distribution of agents' initial opinions.

\begin{theorem}\label{T3}
Under Assumption \ref{A2} and \ref{A3}, system (\ref{FJs}), (\ref{FJh}) achieves opinion consensus if the following conditions hold:\\
(i) there exists a partially stubborn agent which has directed paths consisting of vertices in $V_{p}\bigcup V_{n}$ to any other partially stubborn agents in $\mathcal{G}(W)$;\\
(ii) $\mathcal{G}(H_{ff}(0))$ is connected;\\
(iii) there exists an edge in $\mathcal{G}(H(0))$ between a partially stubborn agent and a fully stubborn agent.
\end{theorem}
{\bf Proof.} From Assumption \ref{A3} and Lemma \ref{L5}, we have that $\mathcal{G}(\Phi^{\prime}(s))$ contains a spanning tree for any $s\in\mathbb{N}$. Hence, Lemma \ref{L4} implies that $\Phi_{1}(s)$ is SIA and $\mathcal{G}(\Phi_{1}(s))$ contains a spanning tree for any $s\in\mathbb{N}$. Let $\Delta(s)=\Phi(s-1)\Phi(s-2)\ldots\Phi(1)\Phi(0)$, then $x(s)=\Delta(s)x(0)$. From the block form of $\Phi(s)$, we have
\begin{equation*}
 \Delta(s)=\begin{bmatrix}
  \Phi_{1}(s-1)\Phi_{1}(s-2)\ldots\Phi_{1}(0) & \ \ \ 0\\
  \Phi_{2}(s-1)\Phi_{1}(s-2)\ldots\Phi_{1}(0) & \ \ \ 0
  \end{bmatrix}.
\end{equation*}
Since the set of all possible matrices $H(s)$ is finite under equation (\ref{FJh1}), $\Phi(s)=H(s)\Psi$, one has that the set of all possible matrices $\Phi(s)$ is finite, which indicates that the set of all possible matrices $\Phi_{1}(s)$ is finite. Moreover, because $\mathcal{G}(\Phi_{1}(s))$ contains a spanning tree for any $s\in\mathbb{N}$, from Lemma \ref{L4.1}, we have $\Phi_{1}(s-1)\Phi_{1}(s-2)\ldots\Phi_{1}(0)$ is SIA for any $s\in\mathbb{N}$. From Wolfowitz Theorem \citep{Wolf63}, there exists a nonnegative vector $\nu\in\mathbb{R}^{r_{1}+r_{2}}$ such that
\begin{equation*}
  \lim_{s\rightarrow\infty}\Phi_{1}(s-1)\Phi_{1}(s-2)\ldots\Phi_{1}(0)=\mathbf{1}_{r_{1}+r_{2}}\nu^{T}.
\end{equation*}
Thus,
\begin{equation*}
\begin{split}
&\lim_{s\rightarrow\infty}\begin{bmatrix}
  \Phi_{1}(s-1)\Phi_{1}(s-2)\ldots\Phi_{1}(0) & \ \ \ 0\\
  \Phi_{2}(s-1)\Phi_{1}(s-2)\ldots\Phi_{1}(0) & \ \ \ 0
  \end{bmatrix}\\
 &=\begin{bmatrix}
  \mathbf{1}_{r_{1}+r_{2}}\nu^{T} & \ \ \ 0\\
  \lim_{s\rightarrow\infty}\Phi_{2}(s-1)\mathbf{1}_{r_{1}+r_{2}}\nu^{T} & \ \ \ 0
  \end{bmatrix}.
\end{split}
\end{equation*}
Since $\Phi_{2}$ is stochastic, we obtain
\begin{equation*}
\lim_{s\rightarrow\infty}\Delta(s)=\begin{bmatrix}
  \mathbf{1}_{r_{1}+r_{2}}\nu^{T} & \ \ \ 0\\
  \mathbf{1}_{n-r_{1}-r_{2}}\nu^{T} & \ \ \ 0
  \end{bmatrix}.
\end{equation*}
Therefore,
\begin{equation*}
 \lim_{s\rightarrow\infty}\lim_{k\rightarrow\infty}x(s,k)=\lim_{s\rightarrow\infty}x(s)=[\nu^{T} \ \mathbf{0}_{n-r_{1}-r_{2}}^{T}]x(0)\mathbf{1}_{n},
\end{equation*}
which implies that system (\ref{FJs}), (\ref{FJh}) achieves consensus as $k\rightarrow\infty, s\rightarrow\infty$. $\blacksquare$

As we discussed in Rmark \ref{R4}, cognitive freezing emerges in model (\ref{FJs}), (\ref{FJs1}) under Assumption \ref{A2}. This is attributed to the externality resulted from the interpersonal influence. Thus, in model (\ref{FJs}), (\ref{FJs1}), fully stubborn agents will keep their opinions unchangeable for all issues. Theorem \ref{T3} shows that the cognitive freezing is weakened by the bounded confidence of agents so that consensus can be achieved even in the presence of fully stubborn agents. Theorem \ref{T2} and Theorem \ref{T3} show that consensus can be achieved by the F-J model over a sequence of issues. Yet in the same context, it generally fails to achieve consensus over a single issue, even under completely connected networks. In practice, as the number of the issues that individuals collaborate to deal with increases, initial opinions of individuals for each issue become more and more similar and reach agreement ultimately. This phenomenon is named cohort effect \citep{Joshi10}, which describes a high degree of similarity in characteristics or beliefs of a group of individuals who share a common set of experiences. Moreover, this mechanism is usually used to train groups for specific tasks (to name but a few, athlete team, special action force or fire brigade) to make the opinions or decisions of the group members grow similar.
\section{\bf Simulations }\label{s-Simulations}
In this section, numerical simulations are provided to illustrate the effectiveness of the theoretical results proposed above.

\begin{example}\label{e1}
(Convergence of the F-J model over a single issue) Consider $10$ agents labeled from 1 to 10 interacting with each other in digraph $\mathcal{G}(W^{1})$ (Fig. \ref{subfig1.1}). $W^{1}$ is the stochastic adjacent matrix of $\mathcal{G}(W^{1})$. Let $\xi^{1}=(0, 0, 0, 0.2, 0.5, 0.7, 1, 1, 1, 1)^{T}$, namely, $V_{f}=\{1, 2, 3\}, V_{p}=\{4, 5, 6\}$ and $V_{n}=\{7, 8, 9, 10\}$. Initialize $x(k)$ with $x(0)=(-1, 0, 1, 1, -2, 0, -1, -2, 1, 2)^{T}$, the trajectory of $x(k)$ under $\mathcal{G}(W^{1})$ with $k$ increasing is shown in Fig. \ref{subfig1.2}.

\begin{figure}
\centering
 \subfloat[t][. $\mathcal{G}(W^{1})$]{
\label{subfig1.1}
\begin{minipage}[t]{0.4\textwidth}
 \centering
  \includegraphics[width=\hsize]{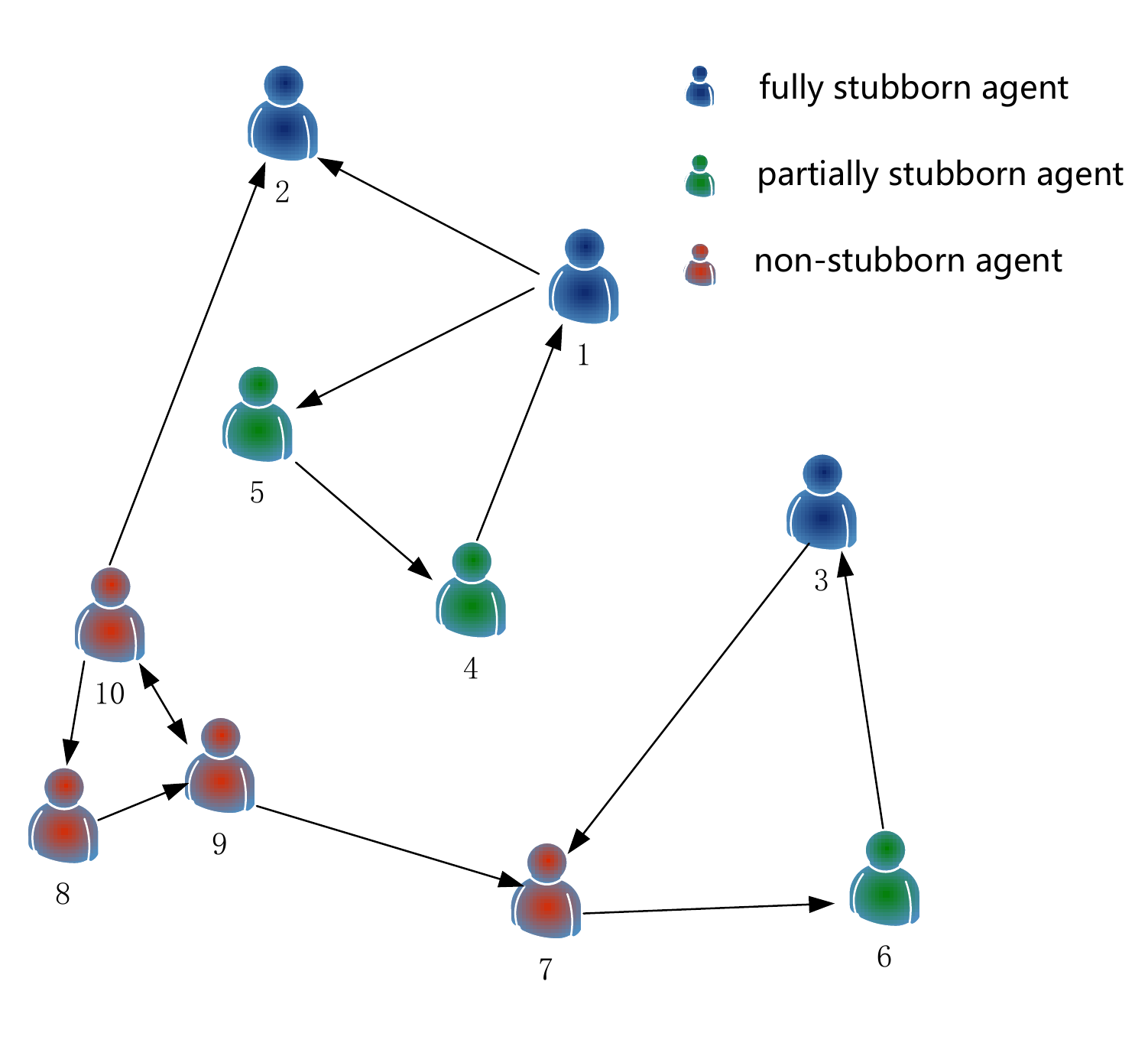}\\
 \end{minipage}
}
\hspace{0pt}
\subfloat[t][. $x(k)$]{
\label{subfig1.2}
\begin{minipage}[t]{0.5\textwidth}
 \centering
  \includegraphics[width=\hsize]{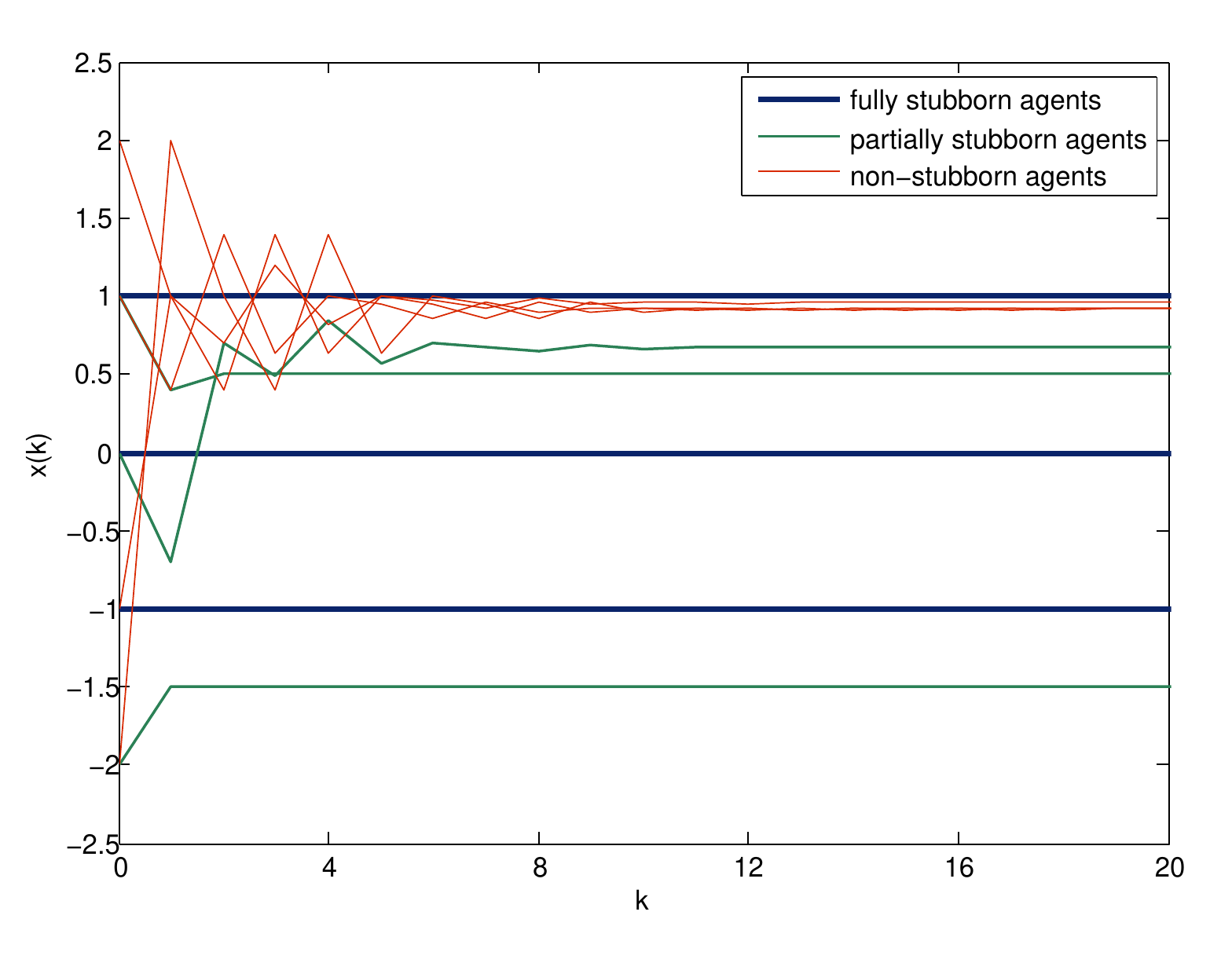}\\
  \end{minipage}
}
\caption{The network topology $\mathcal{G}(W^{1})$ and the trajectory of $x(k)$.}
\end{figure}

Note that in Fig. \ref{subfig1.2}, the opinions of fully stubborn agents are fixed, while the opinions of partially stubborn agents and non-stubborn agents tend to be constant as $k$ increases. The network $\mathcal{G}(W^{1})$ has four SCCs, i.e., $\{1, 4, 5\}$, $\{3, 6, 7\}$, $\{8, 9, 10\}$, $\{2\}$. The ISCC consisting of non-stubborn agents is $\{8, 9, 10\}$, in which agent 9 and agent 10 both have two cycles whose lengths are 2 and 3, respectively. Thus, Assumption \ref{A1} is satisfied. By computing we find that the maximum-modulus eigenvalues of $W^{1}$ are $1, -\frac{1}{2}+\frac{\sqrt{3}}{2}i$, and $-\frac{1}{2}-\frac{\sqrt{3}}{2}i$, while the maximum-modulus eigenvalue of $\Xi^{1}W^{1}$ is $1$, which accords with the algebraic condition in Lemma \ref{L1}. In Fig. \ref{subfig1.2}, the trajectory of $x(k)$ is almost fixed after $k=20$. When $k=20$, one can find that the ISCC $\{8, 9, 10\}$ achieve consensus, while the other non-stubborn agent 7 has different ultimate opinion with them. Moreover, we have $\psi_{7}=0$, $\psi_{8}$, $\psi_{9}$ and $\psi_{10}$ are nonzero, which is consistent with Property \ref{P2} and the discussion following it.
\end{example}

\begin{example}\label{e2}
(Consensus of the F-J model over issue sequences) Consider $10$ agents labeled from 1 to 10 in which agents $1$ to $5$ are partially stubborn agents, others are non-stubborn agents. The network topology is characterized by the digraph $\mathcal{G}(W^{2})$ (Fig. \ref{subfig2.1}). Let $\xi^{2}=(0.3, 0.6, 0.2, 0.8, 0.7, 1, 1, 1, 1, 1)^{T}$. $x(0,0)=(-1, 0, 1, 1, -2, 0, -1, -2, 1, 2)^{T}$ is the initial opinion vector at $s=0$ and $k=0$. The trajectory of $x(k)$ is showed in Fig. \ref{fig3}.
\begin{figure}
\centering
 \subfloat[t][. $\mathcal{G}(W^{2})$]{
\label{subfig2.1}
\begin{minipage}[t]{0.4\textwidth}
 \centering
  \includegraphics[width=\hsize]{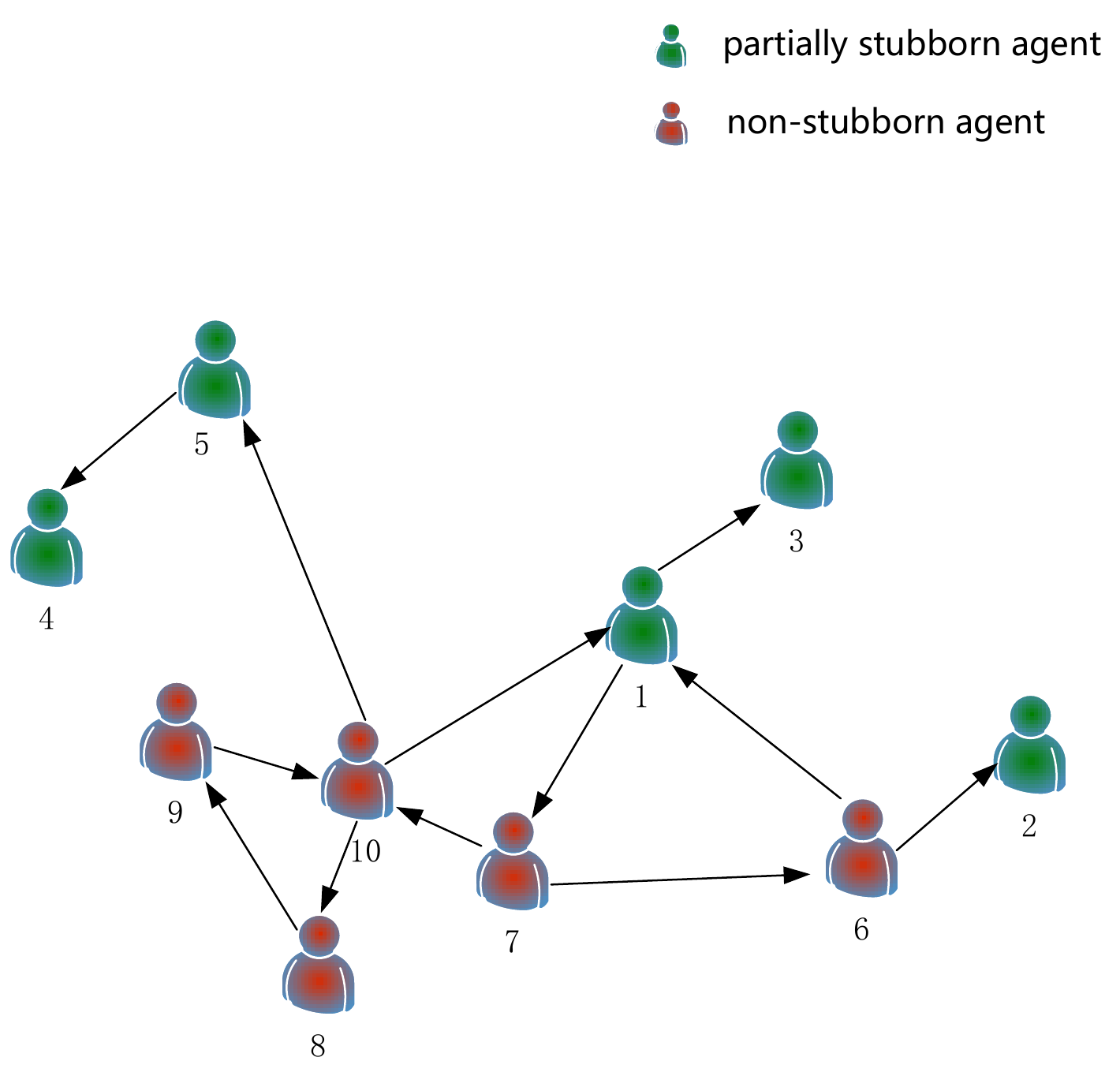}\\
 \end{minipage}
}
\hspace{0pt}
\subfloat[t][. $\mathcal{G}(\Psi^{2})$]{
\label{subfig2.2}
\begin{minipage}[t]{0.4\textwidth}
 \centering
  \includegraphics[width=\hsize]{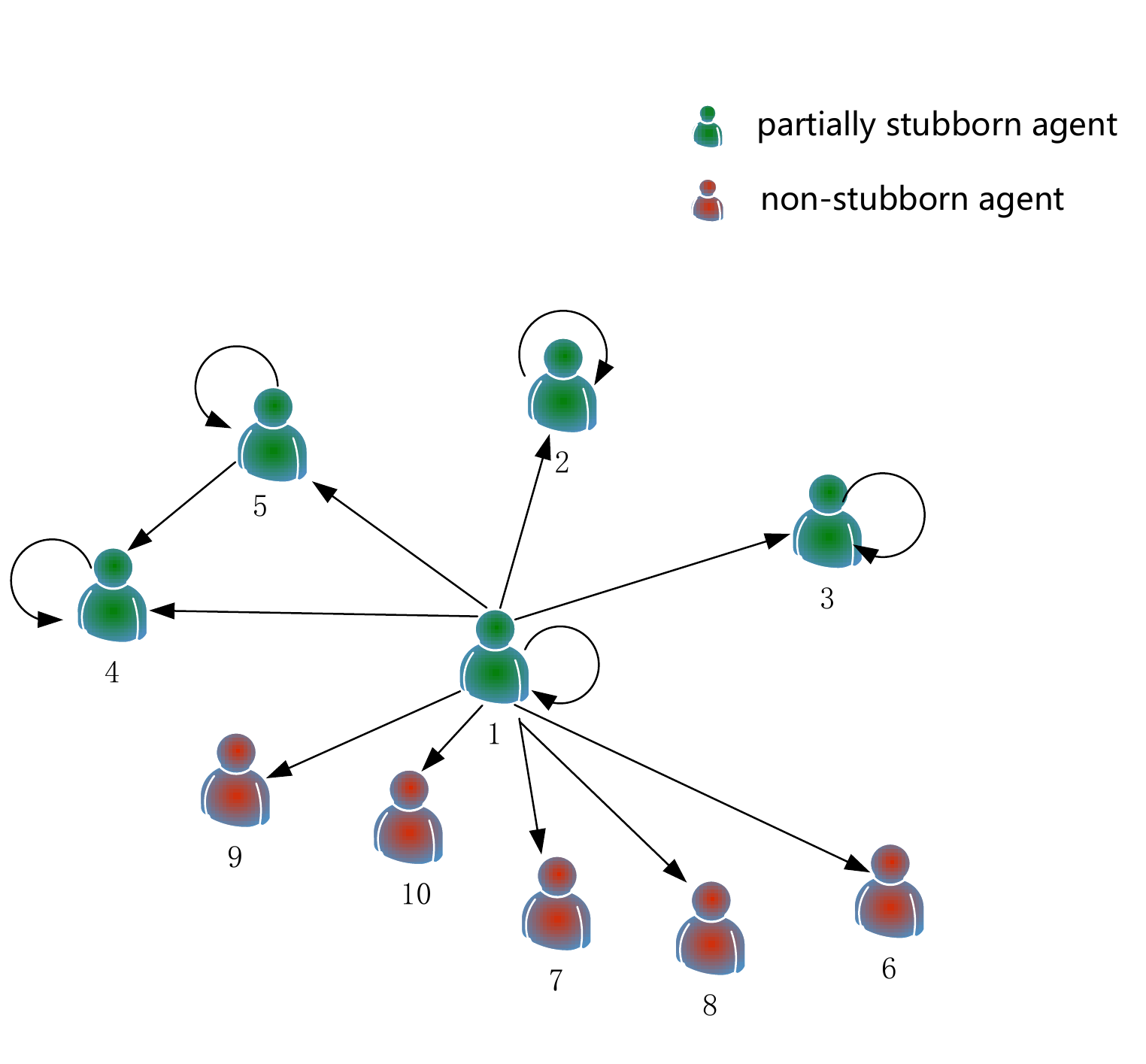}\\
  \end{minipage}
}
\caption{The network topologies $\mathcal{G}(W^{2})$ and $\mathcal{G}(\Psi^{2})$.}
\end{figure}
\begin{figure}
 \centering
  \includegraphics[width=10cm]{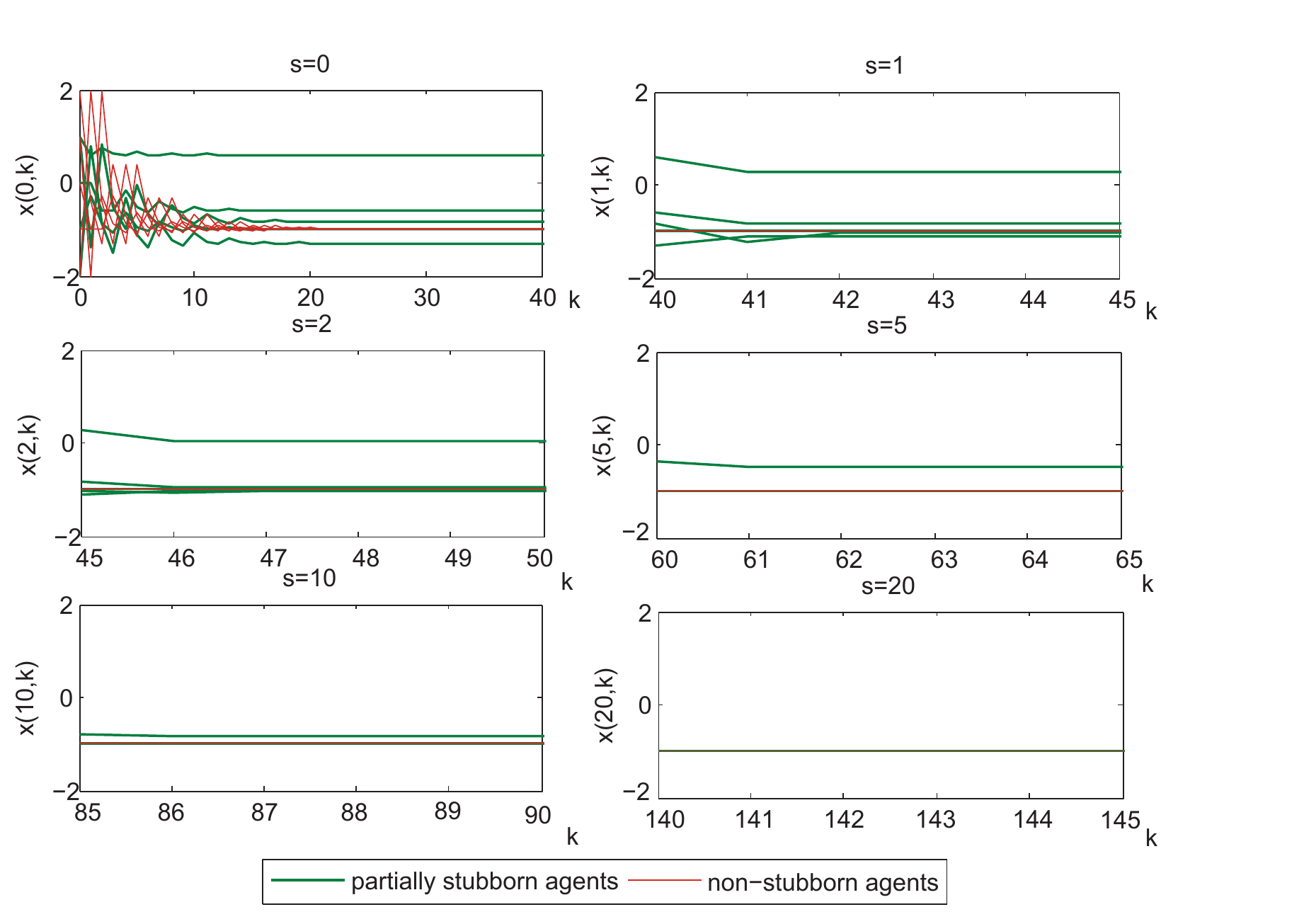}\\
  \caption{The trajectories of $x(s,k)$ at $s=0, 1, 2, 3, 4, 30$, respectively.}\label{fig3}
\end{figure}

Fig. \ref{fig3} shows the trajectories of $x(s,k)$ over issues $s=0, 1, 2, 5, 10, 20$, respectively. Note that over each issue $s$, $x(s,k)$ tends to a steady opinion vector as $k$ increases, and all steady opinions tend to be a common value with $s$ increasing. When $s=30$, Fig. \ref{fig3} shows that consensus is achieved by system (\ref{FJs}), (\ref{FJs1}). Moreover, since $\mathcal{G}(W^{2})$ satisfies the condition in Assumption \ref{A2}, we have $\Psi^{3}=(I-\Xi^{2}W^{2})^{-1}(I-\Xi^{2})$. From Lemma \ref{L3} and Property \ref{P3}, the existence of a partially stubborn agents in $\mathcal{G}(W^{2})$ which has a directed path to other partially stubborn agents implies that the subgraph corresponding to partially stubborn agents in $\mathcal{G}(\Psi^{2})$ not only has a spanning tree, but also contains a star topology, and self-loops are contained by partially stubborn agents in $\mathcal{G}(\Psi^{2})$, which are shown in Fig. \ref{subfig2.2}. We can find visually from Fig. \ref{subfig2.2} that partially stubborn agents do not receive information from non-stubborn agents, which indicates that $\psi_{i}=0$ for any $i\in\{5, 6,\ldots, 10\}$. This is consistent with the theoretical result in Property \ref{P2}.
\end{example}

\begin{example}\label{e3}
(Consensus of the F-J model over issue sequences with bounded confidence) Consider system (\ref{FJs}), (\ref{FJh}) with $10$ agents labeled from $1$ to $10$. The network topology $\mathcal{G}(W^{3})$ is showed in Fig. \ref{subfig3.1}. Let $\xi^{3}=(0, 0, 0, 0.8, 0.3, 0.2, 1, 1, 1, 1)^{T}$, namely, $V_{f}=\{1, 2, 3\}, V_{p}=\{4, 5, 6\}$ and others are non-stubborn agents. Let $x(0,0)=(-0.7, 0.2, 0, 0.2, 2, -2.5, -1.5, 1, 1.5, -1)^{T}$. The confidence bound is set as $d=1$. Since $\frac{1}{2n}<h<\frac{1}{n-1}$, let $h=0.1$. Fig. \ref{subfig4.1} and Fig. \ref{subfig4.2} show the trajectories of $x(s,k)$ and $x(s,0)$, respectively.

\begin{figure}
\centering
 \subfloat[t][. $\mathcal{G}(W^{3})$]{
\label{subfig3.1}
\begin{minipage}[t]{0.4\textwidth}
 \centering
  \includegraphics[width=\hsize]{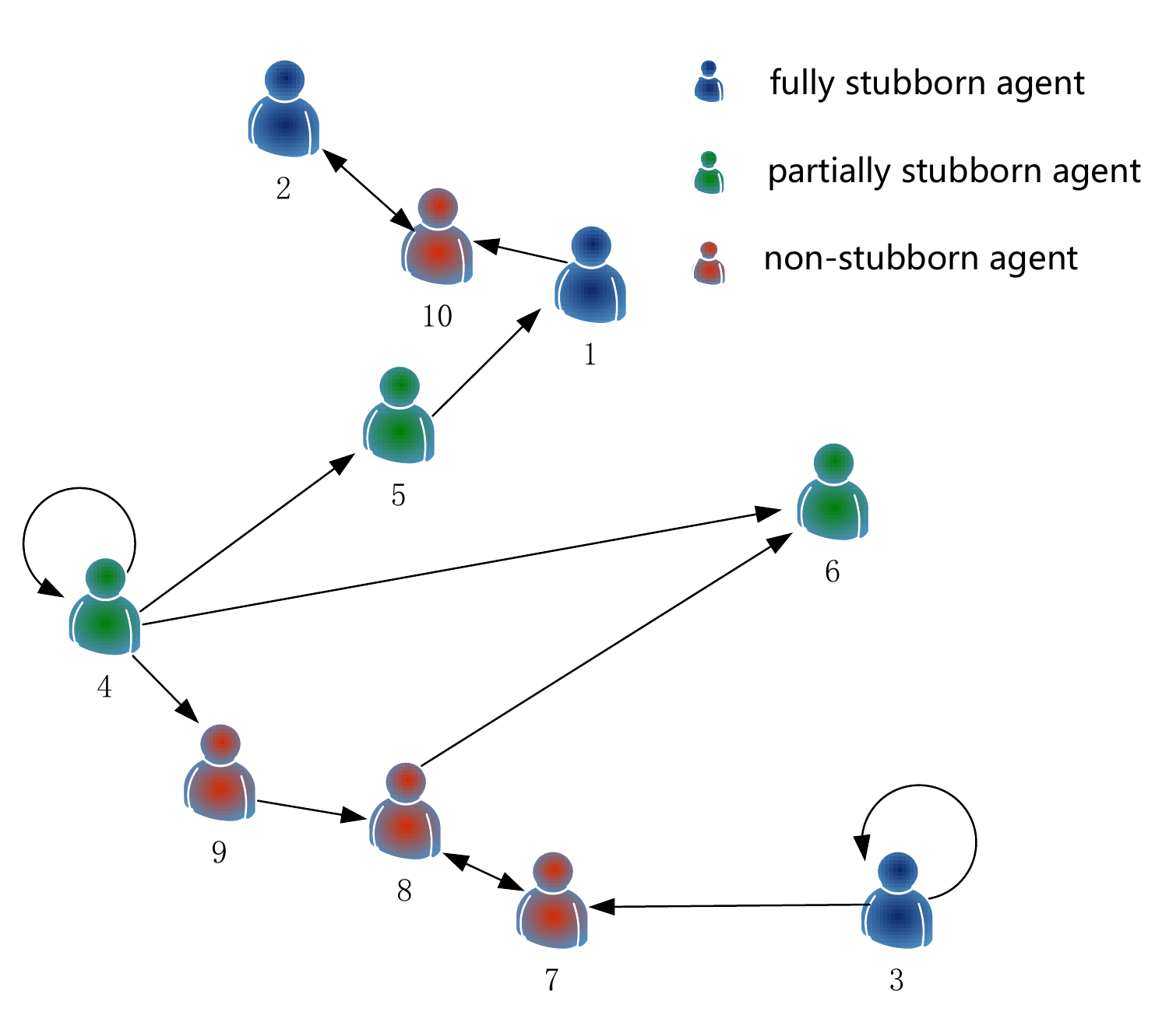}\\
 \end{minipage}
}
\hspace{0pt}
\subfloat[t][. $\mathcal{G}(\Psi^{3})$]{
\label{subfig3.2}
\begin{minipage}[t]{0.4\textwidth}
 \centering
  \includegraphics[width=\hsize]{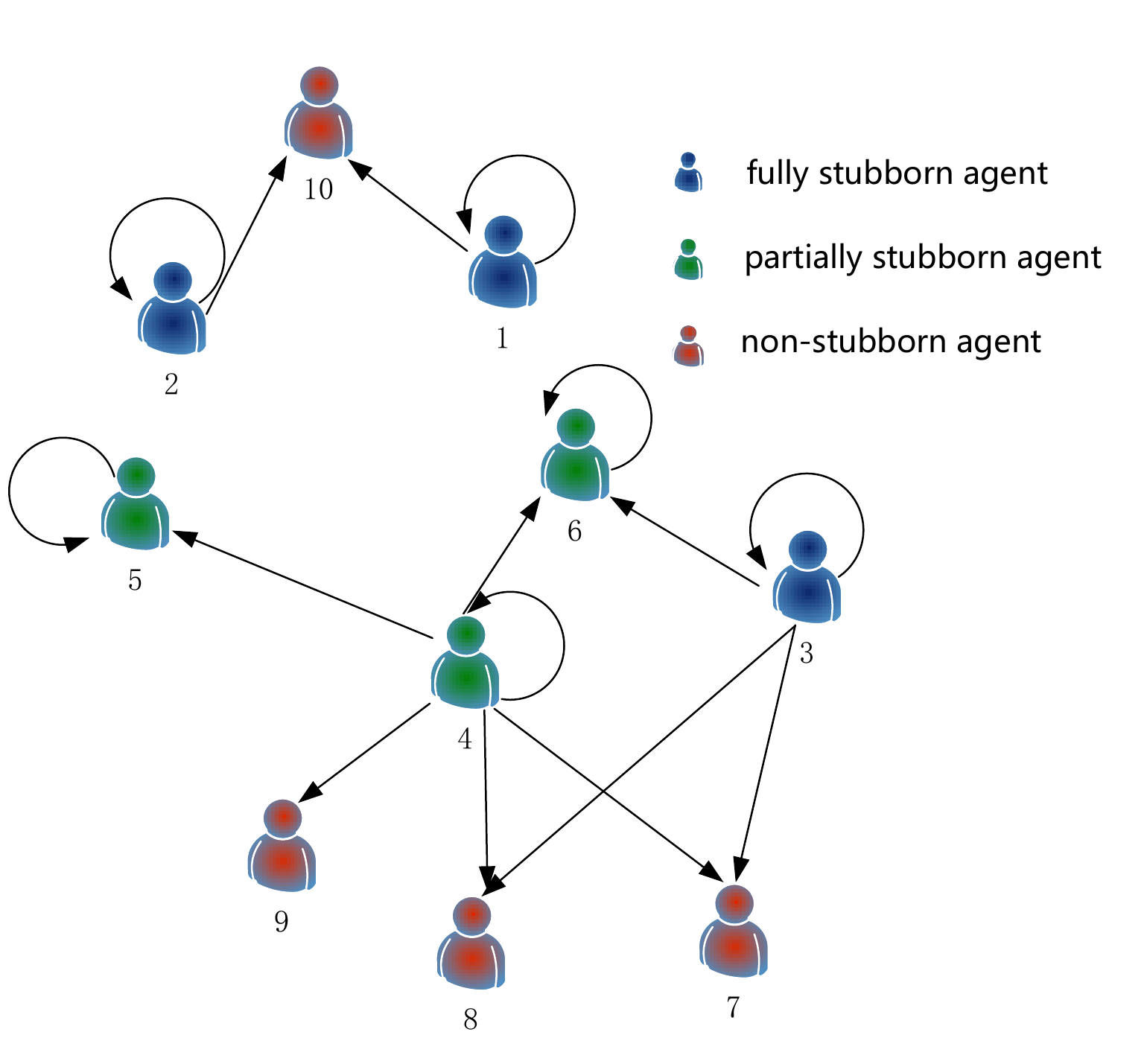}\\
  \end{minipage}
}
\caption{The network topologies $\mathcal{G}(W^{3})$ and $\mathcal{G}(\Psi^{3})$.}
\end{figure}
Because that in Fig. \ref{subfig3.1}, $\mathcal{G}(W^{3})$ satisfies the condition of Assumption \ref{A2}, $\Psi^{3}$ exists. $\mathcal{G}(\Psi^{3})$ is depicted in Fig. \ref{subfig3.2}. Moreover, $\mathcal{G}(\Psi^{3})$ and $x(0,0)$ satisfy the conditions in Assumption \ref{A3} and Theorem \ref{T3}, thus $x(s,0)$ reaches consensus as $s\rightarrow\infty$ (Fig. \ref{subfig4.2}), which implies $x(s,k)$ reaches consensus (Fig. \ref{subfig4.1}). The connectivity of $\mathcal{G}(H(0))$ is related with the percentage of the fully stubborn agents. In this example, Assumption \ref{A3} requires that each fully stubborn agent has at least $8$ neighbors in $\mathcal{G}(H(0))$. By computing we have $y(0)=(-0.7, 0.2, 0, 0.2, 1.46, -1.9876, 0.0082, 0.0274, 0.2$, $ -0.25)^{T}$, thus agents $1, 2, 3, 4, 7$, $8, 9, 10$ form a completely connected subgraph in $\mathcal{G}(H(0))$, and agents $5$ and $6$ are isolated with others, respectively.
\begin{figure}
\centering
 \subfloat[t][. The trajectories of $x(s,k)$.]{
\label{subfig4.1}
\begin{minipage}[t]{0.5\textwidth}
 \centering
  \includegraphics[width=\hsize]{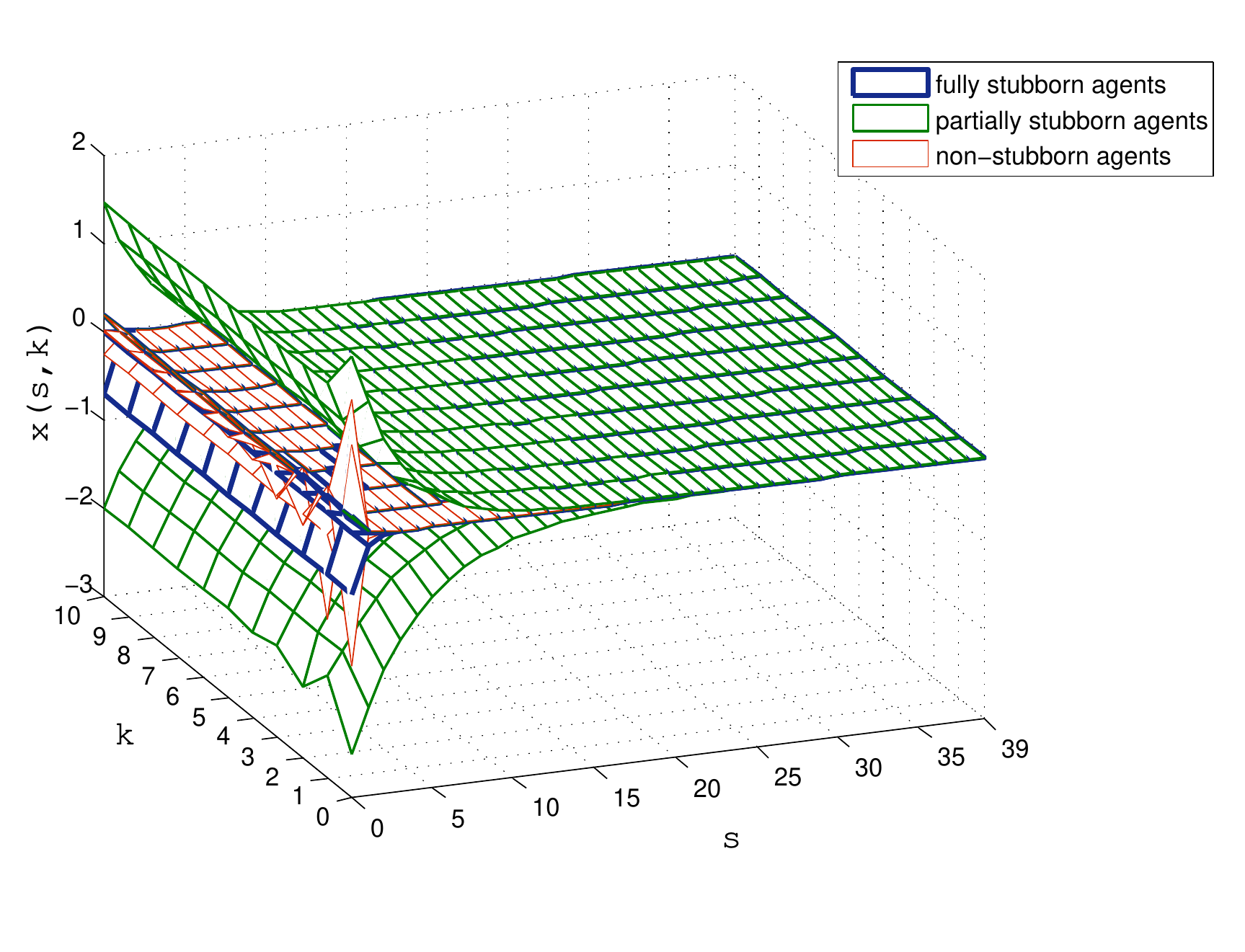}\\
 \end{minipage}
}
\hspace{0pt}
\subfloat[t][. The trajectory of $x(s,0)$.]{
\label{subfig4.2}
\begin{minipage}[t]{0.4\textwidth}
 \centering
  \includegraphics[width=\hsize]{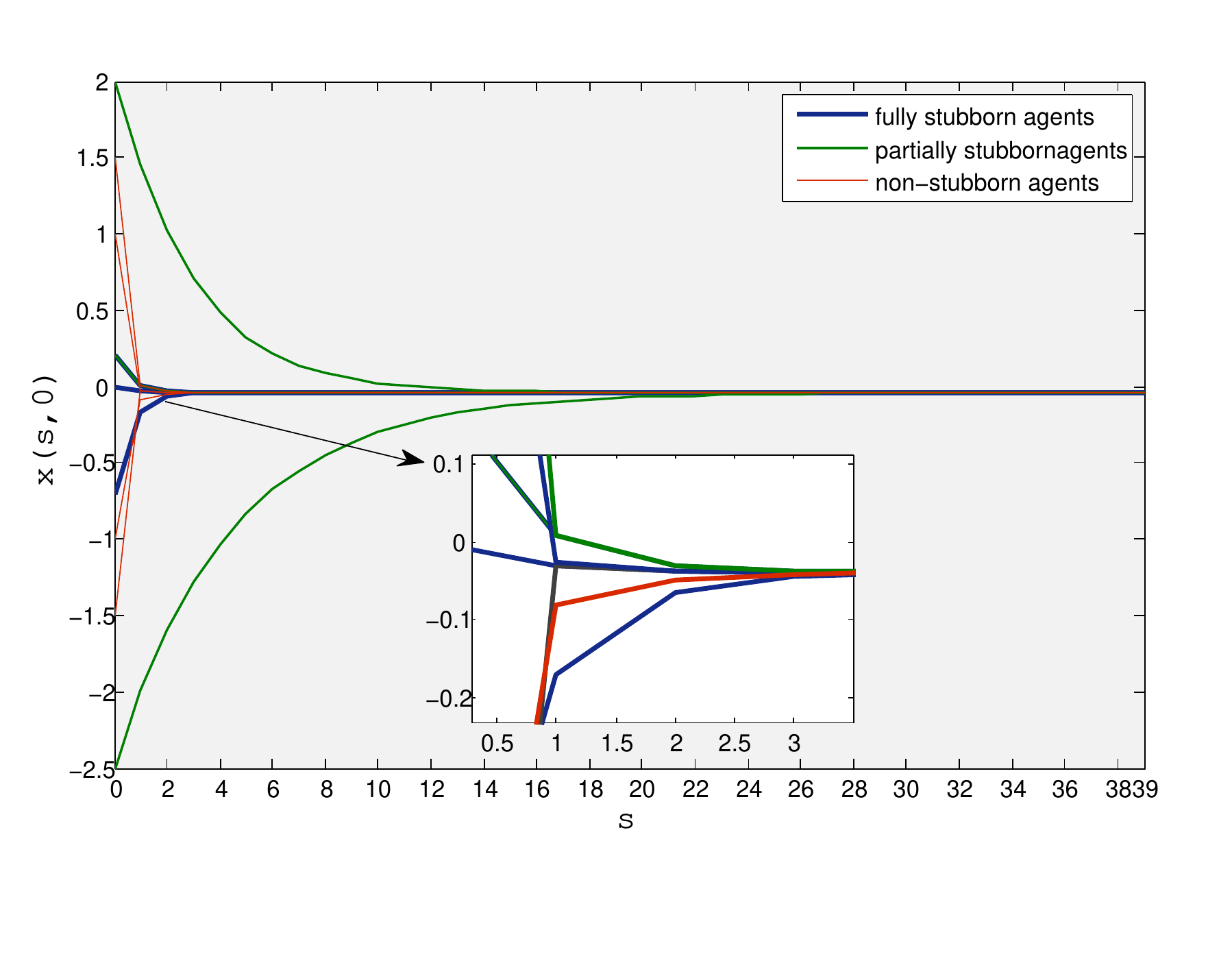}\\
  \end{minipage}
}
\caption{The trajectories of $x(s,k)$ and $x(s,0)$.}
\end{figure}
\end{example}

\section{\bf Conclusions}\label{s-Conclusion}

In this paper, opinion dynamics in social networks with stubborn agents has been investigated over issue sequences. The social network with stubborn agents is described by the F-J model. In order to investigate opinion formation of the F-J model over issue sequences, we studied convergence of the F-J model over a single issue and properties of the ultimate opinion. Then taking advantage of the underlying connections between the network of interpersonal influence and the network describing the relationship of agents' initial opinion for successive issues, opinion consensus and cluster of the F-J model was investigated over issue sequences based on the path-dependence theory. Moreover, we further considered the more general scenario where agents have bounded confidence to others when they form their initial opinions. Mathematical conditions for opinions achieving consensus or forming clusters were established, respectively. We showed that the F-J model may achieve opinion consensus over each path-dependent issue sequence, while it generally does not achieve consensus over a single issue. The connections between network characterizing interpersonal influence and the one describing the relationship of agents' initial opinions for successive issues uncover the difference between opinion formation over issue sequences and over a single issue, such as the enhanced network connectivity in the former. Our investigation may provide some theoretical explanations for many observations in the real world such as decision making in human society or animal kingdom. The future work will focus on the evolution of agents' extent of stubborn over issue sequences.


\end{document}